\documentclass[preprint,12pt]{elsarticle}

\usepackage{amssymb}
\usepackage{latexsym}
\usepackage{mathrsfs}
\usepackage{multirow}
\usepackage{graphicx}
\usepackage{subfig}
\usepackage{caption}
\usepackage{subcaption}
\usepackage{amsmath}
\usepackage{amsthm}
\usepackage{commath}
\usepackage{booktabs}
\usepackage{makecell}
\usepackage[capitalise]{cleveref}
\usepackage{xcolor}
\usepackage{longtable}
\usepackage{booktabs}
\usepackage{float}

\definecolor{mygreen}{RGB}{255,165,0}

\journal{Renewable Energy}

\begin{document}

\begin{frontmatter}



\title{Data-driven modeling of wind farm wake flow based on multi-scale feature recognition}


\author[a,b]{Dong Xu\fnref{fn1}}
\author[c,d]{Zhaobin Li\fnref{fn1}}
\author[c,d]{Xiaolei Yang}
\author[e]{Peng Hou}
\author[f]{Bruno Carmo\corref{cor1}}
\cortext[cor1]{Corresponding author.}\ead{bruno.carmo@usp.br}
\author[g,h]{Xuerui Mao\corref{cor2}}
\cortext[cor2]{Corresponding author.}\ead{maoxuerui@sina.com}
\fntext[fn1]{These two authors contribute equally.}

\affiliation[a]{organization={School of Mechatronical Engineering, Beijing Institute of Technology},
            city={Beijing},
            postcode={100081}, 
            country={China}}

\affiliation[b]{organization={Advanced Research Institute of Multidisciplinary Sciences, Beijing Institute of Technology},
            city={Beijing},
            postcode={100081}, 
            country={China}}

\affiliation[c]{organization={The State Key Laboratory of Nonlinear Mechanics, Institute of Mechanics, Chinese Academy of Sciences},
            city={Beijing},
            postcode={100190}, 
            country={China}}
            
\affiliation[d]{organization={School of Engineering Sciences, University of Chinese Academy of Sciences},
            city={Beijing},
            postcode={100049}, 
            country={China}}

\affiliation[e]{organization={Baima Lake Laboratory Co.,Ltd.},
            city={Hangzhou},
            postcode={310000}, 
            country={China}}

\affiliation[f]{organization={Department of Mechanical Engineering, University of São Paulo},
            city={São Paulo},
            postcode={05508-030}, 
            country={Brazil}}
            
\affiliation[g]{organization={Beijing Institute of Technology (Zhuhai)},
            city={Zhuhai},
            postcode={519088}, 
            country={China}}
\affiliation[h]{organization={State Key Laboratory of Explosion Science and Safety Protection},
            city={Beijing},
            postcode={100081}, 
            country={China}}

\begin{abstract}
Accurate and efficient predictions of wind flow developments with wake effects accounted are crucial for wind farm layouts and power forecasting. Existing methods can be broadly classified as physical measurement, numerical simulations, physics-based modeling, and data-driven modeling. The first two is of high cost in terms of time and resources, the third suffers from low accuracy due to limited physics modeled, while the last one takes advantage of the large amount of high-quality data available and has become increasingly popular. This study proposes a rapid data-driven modeling method for wind farm wake flow, inspired by video frame interpolation and based on the principle of similarity, which utilizes a multi-scale feature recognition technique. The method transforms wind farm field data into images and predicts wake flow by identifying, matching, and interpolating features from a limited set of wake flow images using the Scale-Invariant Feature Transform (SIFT) and Dynamic Time Warping (DTW) approaches. To demonstrate the effectiveness of the proposed method, six representative cases were evaluated, encompassing mini wind farms with varying turbine spacings, different turbine sizes, combinations of spacing and size variations, different numbers of turbines, and various degrees of wind direction misalignment. A Mean Absolute Percentage Error (MAPE) ranging from 0.68\% to 2.28\% is achieved. Due to its ability to flexibly compute both 2D and 3D wake flow fields, the proposed method offers unique computational efficiency advantages over Large Eddy Simulation (LES) and Meteodyn WT in scenarios where two-dimensional wake flow fields are sufficient to meet industrial requirements. Therefore, this method can be employed for the extension of the wake flow database serving wind farm design, power prediction, etc., as an alternative to measurements, numerical simulation, and physics-based modeling, balancing efficiency and accuracy.

\end{abstract}


\end{frontmatter}

\section{Introduction}
In a wind farm, downstream turbines are immersed in the wakes of upstream ones, leading to complex turbine interaction effects, \textit{e.g.} reduced wind speed and power, increased turbulence intensity and fatigue loading, etc. \cite{reddy2020wind}. Therefore, fast and accurate prediction of wake flow in wind farms has been a critical concern for the wind energy community. Methods for such predictions include measurement, numerical simulation, physics-based modeling, and data-driven modeling. 

Various tools such as meteorological towers, LiDAR, Doppler radar, and sodar have been employed for wind turbine wake measurements. Duckworth and Barthelmie \cite{duckworth2008investigation} validated multiple wake models using field data. Kasler et al. \cite{kasler2010wake} used long-range Doppler LiDAR to observe wake recovery downstream of a 5 MW turbine. Bartl et al. \cite{bartl2012wake} investigated wake interactions between two turbines in wind tunnels. Iungo et al. \cite{iungo2013field} measured peak turbulence at the blade tip height using field LiDAR, indicating potential fatigue issues for downstream turbines. Gao et al. \cite{gao2020investigation} developed a 3D Jensen-Gaussian wake model considering wind shear, validated through dual LiDAR experiments. However, the installation and maintenance costs of measurement equipment are notably high, especially in complex terrains and harsh environments such as offshore. Moreover, physical measurement is subject to limitations imposed by weather conditions and equipment performance, and collected data are commonly sparse in space and time.

The physics-based method constructs wake models based on the conservation laws of fluid mechanics and inevitably introduces assumptions to obtain simplified models and their analytical solutions. This method is generally low-cost but has limited accuracy due to the simplified physics assumptions. The Jensen model \cite{jensen1983note} is widely employed to predict wind turbine wakes, assuming axisymmetric expansion described by a distance-based functional formulation. Larsen \cite{larsen1988simple} developed a semi-analytical model leveraging asymptotic expressions derived from Prandtl’s turbulent boundary layer equations. Frandsen’s model \cite{frandsen2006analytical} estimates wind speed deficits in large offshore wind farms with rectangular layouts and uniform turbine spacing. Recent advances include two Gaussian-like models: one by Bastankhah and Porté-Agel \cite{bastankhah2014new}, which assumes a Gaussian distribution for the velocity deficit in the wake, and the other by Xie and Archer \cite{xie2015self}, which incorporates anisotropic wake expansion. Ghaisas et al. \cite{ghaisas2016geometry} introduced a hybrid approach based on the geometric properties of the wind farm.

The numerical simulation of wind farm wake flows mainly involves Computational Fluid Dynamics (CFD), which solves the Navier–Stokes equations. Common CFD approaches include Reynolds-Averaged Navier–Stokes (RANS) and Large Eddy Simulation (LES). RANS has been widely used for steady-state wind farm wake simulations by time-averaging turbulence effects. Castellani et al. \cite{castellani2013application} applied actuator disc models within RANS for coastal wind farms, improving prediction accuracy. Van der Laan et al. \cite{van2015improved} enhanced near-wake predictions by incorporating tip vortex corrections into a $\text{k--}\varepsilon$ RANS model. Unlike RANS, which parameterizes all turbulent scales, LES explicitly resolves large eddies while modeling only small sub-grid scales, rendering it more suitable for detailed wake studies.  Wu and Porté-Agel \cite{wu2015modeling} demonstrated that LES with dynamic actuator disk models can accurately predict power losses in offshore wind farms. Zhong et al. \cite{zhong2015lagrangian} captured near-wake vortex structures using Lagrangian dynamic LES combined with actuator line methods, validated against wind tunnel data. Archer et al. \cite{archer2018review} highlighted LES advantages over analytical models for offshore and onshore wakes. He et al. \cite{he2021novel} further proposed a 3D elliptical Gaussian model incorporating wind shear and anisotropic expansion, validated against LES and experimental results for improved far-wake predictions. Besides CFD, the Weather Research and Forecasting (WRF) model is widely used for simulating wind farm flows. Developed by NCAR and NCEP, WRF accounts for atmospheric physical processes and can be nested down to meso- and micro-scales. Udina et al. \cite{udina2020wrf} incorporated WRF-LES to simulate wind farm wakes over complex coastal terrains, demonstrating improved wake recovery predictions.

In numerical simulations of wind farm flow, wind turbines are often simplified using models such as the Actuator Disk Model (ADM), Actuator Line Model (ALM), and Actuator Surface Model (ASM) to reduce computational cost by avoiding detailed blade geometries \cite{wang2019numerical}. In ADM, blade effects are represented by volumetric momentum sources, modeling the turbine as a drag disk. In ALM, the blades are replaced by virtual lines that distribute local lift and drag along the span. Martínez-Tossas and Meneveau \cite{martinez2015large} showed that ALM generates more realistic and localized flow patterns than ADM, resulting in more accurate predictions of near-wake structures and wake deflection near blade tips. The ASM further refines the actuator modeling approach by representing turbine blades as full surfaces, which improves accuracy but also increases computational expense significantly \cite{burton2011wind}.
Despite these modeling advancements, the high computational cost of numerical simulations remains a major limitation, especially for optimization tasks such as turbine design and wind farm layout, even when using simplified models like ADM.

Data-driven methods have been widely applied in wind energy, meteorology, and fluid dynamics due to the availability of high-quality large-scale data. Recent studies have developed hybrid models combining neural networks for offshore wind assessment, predictive control frameworks optimizing power and loads, and machine learning approaches for load forecasting and turbulence model correction \cite{zhu2022review}. Deep learning has also been used for global weather prediction \cite{weyn2020improving} and for enhancing reduced-order models through data-driven closures \cite{xie2018data}. A critical question in data-driven modeling is the construction of the reduced-order model (ROM) or the extraction of essential dynamics or features. Existing methods include Principal Component Analysis (PCA) targeting at low-dimensional data, whose high-dimensional counterpart is Proper Orthogonal Decomposition (POD). Applying the POD-based method to wind farm wake data extracts low-dimensional wake structures, reducing data complexity. Debnath et al. \cite{debnath2017towards} used POD and Dynamic Mode Decomposition (DMD) to study coherent vorticity structures in the wake of a single turbine. Hamilton et al. \cite{hamilton2018generalized} identified dynamic wake modes using POD. DMD-based methods, in addition to providing ROMs, are also used for wake prediction. Ali and Cal \cite{ali2020data} predicted wake flow in a turbine array using a Hankel-based DMD method. Chen et al. \cite{chen2022dynamic} applied extended DMD to forecast the wake of a single turbine and assessed its robustness to noise. Studies have also combined POD and DMD with forecasting techniques for wake flow prediction. Iungo et al. \cite{iungo2015data} integrated DMD with the Kalman filter for single-turbine wake flow prediction. Ali et al. \cite{ali2021clustering} used POD-based clustering to evaluate the optimal sparse sensor locations downstream of a turbine and subsequently applied LSTM to predict the fluctuating velocity. Zhang and Zhao \cite{zhang2020novel} employed POD and LSTM to predict the reduced coefficients of wake. Additionally, spatio-temporal Koopman decomposition \cite{le2019new}, akin to POD and DMD, has been used to analyze turbine wake structures.

Besides those reviewed data-based reduced order modeling, Artificial Intelligence (AI) based data-driven methods, \textit{e.g.}, artificial neural networks (ANNs), decision trees, k-nearest neighbors, linear and polynomial regression, and support vector machines have been flourishing in wake flow prediction. AI methods can process data from various sources, including observations, turbine operations, and simulations. Nai-zhi et al. \cite{nai2022data} used Genetic Algorithms (GA) and Random Forest (RF) to develop a wind farm wake model incorporating Supervisory Control and Data Acquisition (SCADA) and analytical model data. Ti et al. \cite{ti2020wake} combined RANS simulations with Back Propagation Neural Networks (BPNN) for wake prediction. Zhang and Zhao \cite{zhang2022wind} used Generative Adversarial Networks (GANs) and Simulator fOr Wind Farm Applications (SOWFA) simulations to predict wake flow and assess the wake of multiple turbines under varying conditions. Kirby et al. \cite{kirby2023data} applied GPR to predict thrust coefficients using both low-fidelity RANS and high-fidelity LES data.

However, these data-driven methods, especially those based on AI techniques, rely on sufficient and reliable wake flow datasets for model training and validation. The size of data used in the studies mentioned above ranges from dozens to millions of snapshots, depending on the complexity of their tasks as listed in Table \ref{table ref lists}. Insufficient or low-quality data has been reported to impact the accuracy and reliability of the models \cite{zhang2023novel}. Due to the challenges of obtaining large amounts of high-quality flow field data, a method that interpolates high-quality data from a limited dataset will significantly alleviate the current challenges in the wind energy community.

\begin{longtable}{lm{3em}lm{3em}lm{3em}lm{5cm}}
\caption{Summary of works on data-driven wind-farm wake flow modeling.} \\
\toprule
Studys & Input data & Method & Data size \\
\midrule

\endfirsthead

\toprule
Studys & Input data & Method & Data size \\
\midrule

\endhead

\bottomrule
\endfoot

Iungo et al. \cite{iungo2015data}	& LES & DMD & 214 snapshots \\
Debnath et al. \cite{debnath2017towards}	& LES & POD, DMD & 282 and 387 snapshots \\
Ali and Cal \cite{ali2020data}	 & Wind tunnel & DMD & 40 kHz, 4,000,000 data points. \\
Hamilton et al. \cite{hamilton2018generalized}	& LES & POD & 2,000 snapshots \\
Zhang and Zhao \cite{zhang2020novel} & LES & POD-RNN & 710 snapshots \\
Ali et al. \cite{ali2021clustering} & LES & POD-k-means, RNN & 2,000 snapshots \\
Chen et al. \cite{chen2022dynamic} & LES & Extended DMD & 750 snapshots \\
Ti et al. \cite{ti2020wake} & RANS & BPNN & 403 simulations \\
Zhang and Zhao \cite{zhang2022wind} & LES & GAN & 270 samples \\
Renganathan et al. \cite{ashwin2022data} & LiDAR& ANN, GPR & 6,654 scans \\
Nai-zhi et al. \cite{nai2022data} & SCADA & GA, RF & 2,766 points \\
Kirby et al. \cite{kirby2023data} & LES & GPR & 50 dataset \\
\label{table ref lists}
\end{longtable}

In a broader view, modeling of intermediate data from the limited dataset is also a key technique in video frame interpolation, which involves generating intermediate frames from two known video frames to increase the frame rate. Common methods include direct approach, kernel-based approach, phase-based approach, and wrapping-based approach. The wrapping-based approach stands out for its ability to accurately interpolate key features by leveraging local feature extraction and fusion \cite{tulyakov2022time}. 

Inspired by video frame interpolation, we propose a fast wind farm flow field modeling method using limited data. The proposed method transforms wind farm wake field data into images and leverages inherent similarities in wake flow fields to identify, match, and interpolate key features using the Scale-Invariant Feature Transform (SIFT) and Dynamic Time Warping (DTW) techniques. SIFT, a multi-scale feature recognition method, identifies and matches features at different scales, enabling an in-depth understanding of the development of critical features incorporated in the data \cite{adelson1984pyramid}. It has been widely used in image processing, including surface reconstruction, edge detection, and noise reduction. By appropriately selecting scale ranges and feature descriptors, SIFT-based methods can extract rich information from a limited number of images. For example, Wang et al. \cite{wang2020scale} improved matching accuracy in aerial images with optimized scale selection, enhancing performance when few images are available. Li et al. \cite{li2021multi} proposed a multi-scale SIFT approach to robustly extract features under varying resolutions with limited data. Yan et al. \cite{yan2021infrared} enhanced infrared image matching from small datasets by integrating adaptive scale descriptors into the SIFT framework. Therefore, multi-scale feature recognition methods have become an efficient and reliable solution for scenarios with high data acquisition costs or limited annotated data. However, SIFT-based methods have not yet been applied in wind energy or fluid dynamics.

The proposed data-driven wind farm wake modeling method requires only two wake flow field snapshots as input, rendering it one of the methods with the lowest input requirements currently available for wind farm wake flow modeling. Additionally, to the best of our knowledge, this study represents an initial attempt to apply multi-scale feature recognition methods to data generated in dynamic wind farm systems, thereby extending the application scope reported in previous studies.

The rest of the paper is organized as follows. In Section 2, we provide specific details of the proposed data-driven modeling method. The methods and parameter settings for computing wake flow field data with CFD are detailed in Section 3. In Section 4, the accuracy of the designed data-driven method is comprehensively validated through six case studies. In Section 5, conclusions are drawn.

\section{Methodology}

In a wind farm, the primary parameters affecting the wake flow are turbine spacing and size. Meanwhile, wind development in a farm, especially on a large scale, exhibits clear similarities due to the wake effect \cite{bastankhah2014new}. Thus, it is possible to identify key features in sample flows and generate flow fields associated with different wind farm parameters through feature interpolation. The proposed modeling method consists of three main components: wake flow field feature point identification, feature point matching, and feature interpolation. The feature point identification process is akin to the SIFT method, with the key difference that the wake flow field data are converted into streamline data to generate more feature points. Feature point matching is based on descriptors such as scale, label, and position, as well as the similarity of streamlines between feature points, calculated using the DTW method. Feature points are considered successfully matched only when their descriptors are consistent and the mean similarity among all feature point pairs is maximized. 

Upon successful matching of feature points, interpolation algorithms are used to generate new wake flow fields. To evaluate the performance of the proposed method, the Mean Absolute Error (MAE) and Mean Absolute Percentage Error (MAPE) are calculated.

\begin{figure}[H]
\centering
\includegraphics[width=0.9\textwidth]{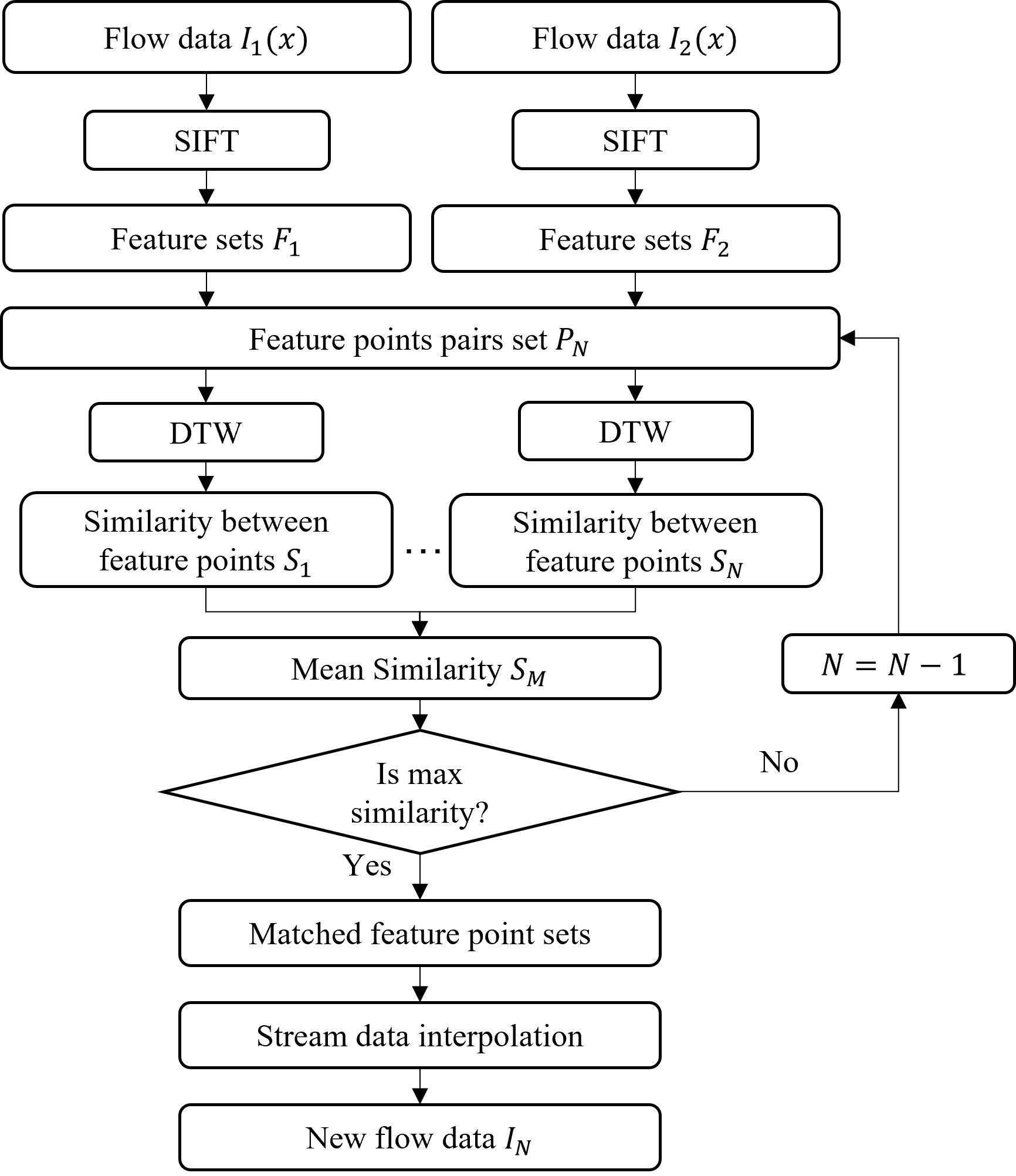}
\caption{Flowchart of proposed data-driven modeling method.}
\label{proposed}
\end{figure}

The workflow of the proposed modeling method is outlined in Figure \ref{proposed}. The key techniques used in this process are elaborated on below. 

\subsection{Scale-Invariant Feature Transform}
\begin{figure}[H]
\centering
\includegraphics[width=0.9\textwidth]{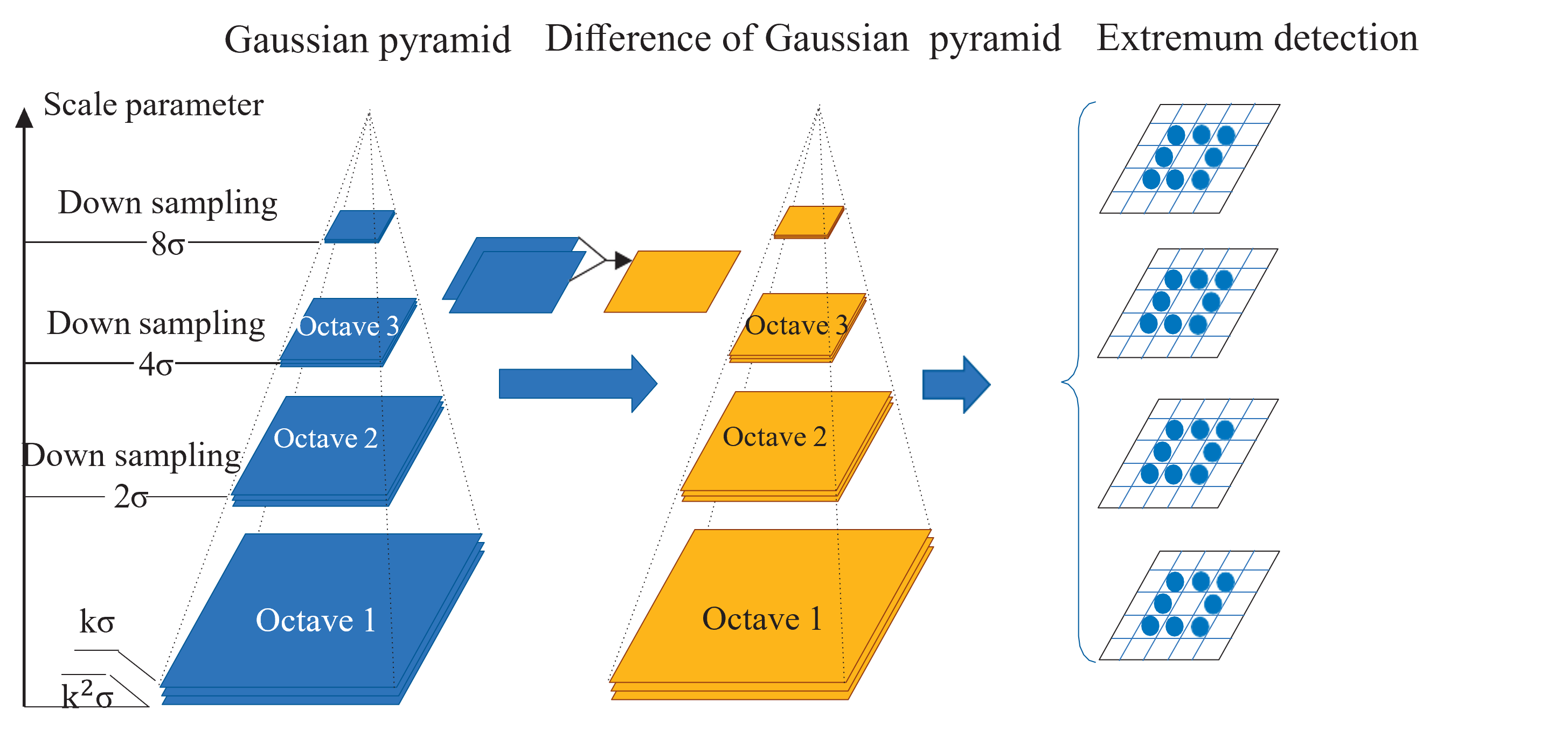}
\caption{Flowchart of SIFT \cite{wang2022scale}.}
\label{SIFT}
\end{figure}

SIFT is a computer vision algorithm for detecting and describing local features in images. It searches for extrema in scale space and computes descriptors that are invariant to translation, scale, and rotation, and it has been widely used in image recognition. The specific process of SIFT matching is shown in Figure \ref{SIFT} and its main steps are listed below.

1) Construction of Gaussian pyramid. 

The field data are first transformed using a Gaussian filter with different scale factors. This filtering is achieved through the convolution of the field data with the Gaussian convolution kernel:
\begin{equation}
\label{eq:L}
L(x, \sigma)=G(x, \sigma)*I(x),
\end{equation}
where $x$ is the coordinate in the sampling space. Different sampling intervals of $x$ construct field data $I(x)$ with various sampling scales. $G(x,\sigma)$ is the convolution kernel function of the Gaussian, defined as:
\begin{equation}
\label{eq:G}
G(x, \sigma)=\frac{1}{2 \pi \sigma} * \exp \left(-\frac{x^{2}}{2 \sigma^{2}}\right),
\end{equation}
where $\sigma$, known as the Gaussian scale factor, represents the variance of the Gaussian distribution and determines the smoothing scale of the flow data. 

Therefore, the transformed field data at different sampling intervals create groups of wake flow field data, with each group containing multi-layer field data obtained through Gaussian filtering. As shown in Figure \ref{SIFT}, at higher sampling intervals of wake flow field data, the scale-spaces $L(x, \sigma)$ decrease. Different scale-spaces $L(x, \sigma)$ form a Gaussian pyramid, with each scale space consisting of a group of flow field data, referred to as an Octave.

2) Extremum detection of the scale spaces.

The difference between two adjacent layers in the Gaussian pyramid, known as the Difference of Gaussian (DoG) pyramid and shown in Figure \ref{SIFT}, can be calculated as: 
\begin{equation}
\label{eq:DOG}
D(x, \sigma)=L(x, k\sigma)-L(x, \sigma),
\end{equation}
where $k$ is the ratio of the scale transformation factor between the adjacent layers. In the present work, $k$ is set to $\sqrt{2}$ for layers within the same group and $2$ in the layers of different groups \cite{wang2022scale}.

Each point in the Difference of Gaussian (DoG) pyramid is compared with its neighbors in both the same layer and adjacent layers to determine if it is an extremum point.

\subsection{Dynamic Time Warping}
\begin{figure}[H]
\centering
\includegraphics[width=0.9\textwidth]{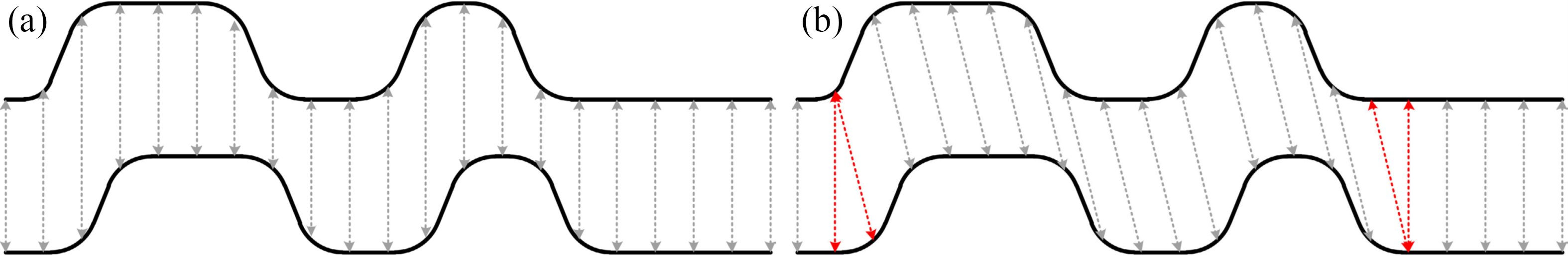}
\caption{Alignment of two time series (the arrows represent aligned points) \cite{choi2020fast}. (a) Euclidean distance, (b) DTW distance.}
\label{DTW}
\end{figure}

The DTW algorithm has been widely adopted to measure the similarity between two data sequences. Unlike the one-to-one correspondence in the Euclidean distance algorithm, DTW warps sequences in a nonlinear fashion to align them (Figure \ref{DTW}). DTW minimizes distortion effects by using an elastic transformation to align similar phases between different patterns. Even with deformations between sequences, DTW identifies the greatest similarities between them  \cite{choi2020fast}. 

DTW calculates the similarity between two sequences by finding the optimal alignment. It computes the distance between points in each sequence and accumulates these distances to measure similarity. These distances are represented by the cost matrix, while the accumulated distances are represented by the accumulated cost matrix. 
The cost matrix $C(m, n)\in R^{d_{1}\times d_{2}}$, $1\leq n \leq d_{1}$, $1\leq m \leq d_{2}$ for the distances between two time series $x=(x_{1}, x_{2}, \ldots, x_{d_{1}-1}, x_{d_{1}})$ and $y=(y_{1}, y_{2}, \ldots, y_{d_{2}-1}, y_{d_{2}})$ is given as follows:
\begin{align}
C(n,m)=(x_n-y_n)^{2}, 
\label{eqn:dtw1}
\end{align}
where, $d_{1}$ and $d_{2}$ are the lengths of time series $x$ and $y$, respectively, and the cost matrix is used to calculate the accumulated cost matrix. 
The optimal warping path $p$, which is a set of index points crucial for determining the optimal alignment, is defined as follows:
\begin{align}
p=(p_{1}, p_{2}, \ldots, p_{l}, \ldots, p_{L-1}, p_{L}) \ {\rm and} \ p_{l}=(n_{l}, m_{l})\in C, l\in[1, L],
\label{eqn:dtw2}
\end{align}
where $L$ is the length of the optimal warping path. The optimal warping path of the accumulated cost matrix is subject to boundary condition, monotonicity condition, and step size condition.

1) Boundary condition

The starting and ending points of the optimal warping path are as follows:
\begin{align}
p_{1}=(1, 1), \ p_{L}=(d_1, d_2).
\label{eqn:dtw3}
\end{align}

2) Monotonicity condition

The subsequent index value of the optimal warping path must be greater than or equal to the current index value.
\begin{align}
n_1\leq n_2 \leq \ldots \leq n_{L-1} \leq n_{L}, \\
m_1\leq m_2 \leq \ldots \leq m_{L-1} \leq m_{L}. 
\label{eqn:dtw4}
\end{align}

3) Step size condition 

The difference between neighboring values in the optimal warping path has a step size defined as follows:
\begin{align}
p_{l+1}-p_{l}\in{(1, 0),(0, 1),(1, 1)},l\in [1, L].
\label{eqn:dtw5}
\end{align}

Accumulated cost matrix $A(n, m) \in R^{d_{1}\times d_{2}}$, $1\leq n \leq d_{1}$, $1\leq m \leq d_{2}$ is calculated as follows:
\begin{align}
A(n, m)=\begin{cases}C(n, m) & {\rm if} \; n=1 \; {\rm and} \; m=1\\ C(n, m)+A(n-1, m) & {\rm if} \; n\geq2 \; {\rm and} \; m=1\\ C(n, m)+A(n, m-1) & {\rm if} \; n=1 \; {\rm and} \; m\geq2\\ C(n, m)+\begin{cases}A(n-1, m-1) & \\ A(n-1, m) & \\ A(n, m-1) & \end{cases} & {\rm if} \; n\geq2 \; {\rm and} \;m\geq2.\end{cases}
\label{eqn:dtw6}
\end{align}

After computing the accumulated cost matrix, the optimal warping path is determined by tracing the smallest values from $A(d_{1}, d_{2})$ to $A(1, 1)$. In this manner, DTW measures the similarity between two sequences, and the DTW distance is expressed as:
\begin{align}
DTW(x, y)=A(d_{1}, d_{2}).
\label{eqn:dtw7}
\end{align}

\begin{figure}[H]
\centering
\includegraphics[width=\textwidth]{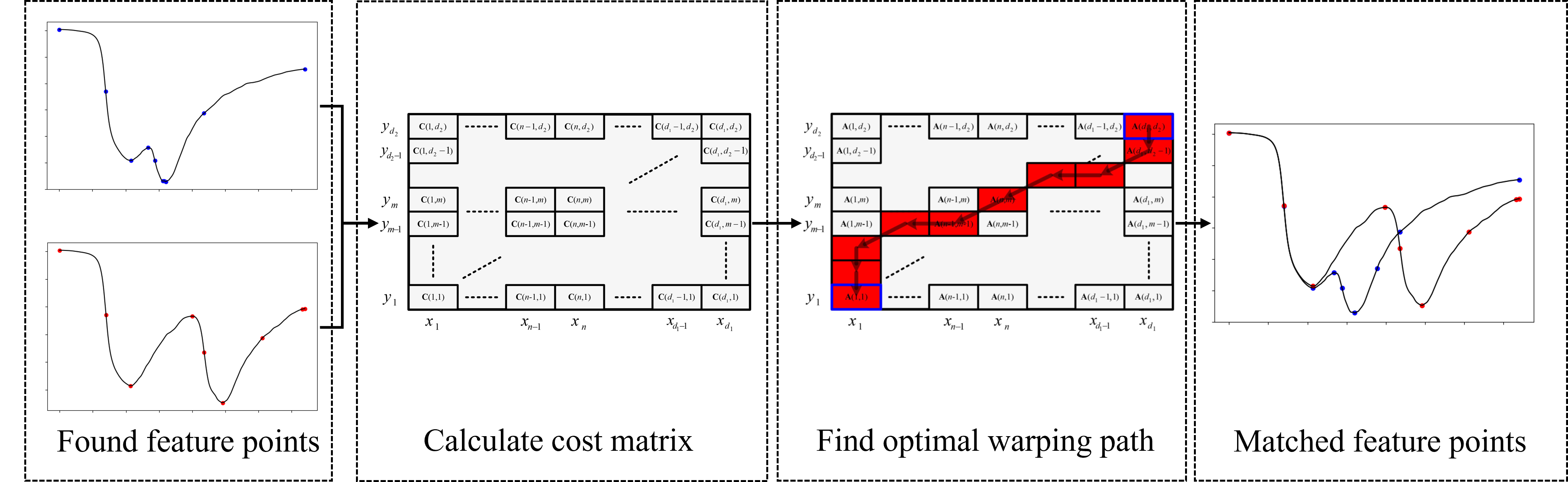}
\caption{Process flowchart for matching feature points in wake flow data using DTW.}
\label{feature_match}
\end{figure}

The process of matching feature points in wake flow field data using DTW is illustrated in Figure \ref{feature_match}.

\subsection{Interpolation Method}

In this study, a curve-based interpolation algorithm is employed. Following the establishment of correspondences between feature points, the curves connecting these feature points are resampled at equal intervals to ensure that the number of data points on both curves remains identical. Subsequently, linear interpolation is conducted between the corresponding points of the two curves to generate the interpolated ones. This interpolation approach has been widely adopted in the field of computer vision, and its detailed formulations can be found in the work of Henry Johan et al.~\cite{johan2000morphing}.

\subsection{Evaluation of indicators}
To evaluate the performance of the proposed method, both qualitative and quantitative measures will be used. The qualitative assessment is provided through error maps, while the quantitative evaluation employs two metrics to assess interpolation performance, \textit{i.e.} MAE and MAPE, defined as follows:
\begin{align}
MAE&=\frac{1}{n}\sum_{i=1}^{n}{\left | U_i - \widetilde{U_i} \right |}, \\ 
MAPE&=\frac{1}{n}\sum_{i=1}^{n}{\left | \frac{U_i - \widetilde{U_i}}{U_i} \right |},
\end{align}
where $U_i$ is the target and $\widetilde{U_i}$ is the interpolated value, and $n$ is the total grid number. 

\section{Data collection method}

\subsection{LES Configuration for Wind Farm Simulations}

\subsubsection{Numerical method}
LES is adopted to predict the unsteady wakes of wind turbines under turbulent inflow conditions mimicing the atmospheric boundary layer. The simulations are carried out using the VFS-Wind code \citep{yang2015VFS}, which solves the filtered incompressible Navier-Stokes equations
\begin{align}
\boldsymbol{\nabla} \cdot {\boldsymbol{u}} & = 0, \label{eqn:cnt} \\ 
\frac{\partial {\boldsymbol{u}}}{\partial t}+({\boldsymbol{u}} \cdot \boldsymbol{\nabla}) {\boldsymbol{u}} & =-\frac{1}{\rho} \boldsymbol{\nabla} {p} + \nu \boldsymbol{\nabla}^2 {\boldsymbol{u}} - \boldsymbol{\nabla} \cdot \boldsymbol{\tau} + \frac{\boldsymbol{f}}{\rho},
	 \label{eqn:ns}
\end{align}
where $\boldsymbol{u} = {\left (u_x, u_y, u_z \right )}^{T}$ represents the velocity vector in Cartesian coordinates, where $x$, $y$, and $z$ represent the streamwise, transverse, and vertical directions, respectively. Here, $p$ denotes the pressure, $\rho$ is the fluid density, $\nu$ indicates the fluid kinematic viscosity, and $\boldsymbol{f}$ is the body force induced by the turbine using the ADM. The subgrid-scale stress, $\boldsymbol{\tau}$, arising from the filtering of the nonlinear convection term is modelled using the dynamic Smagorinsky model. To map these forces onto the background grid nodes and prevent singularities, a smoothed discrete delta function is employed \cite{yang2009Kernel}.

\begin{figure}[H]
\centering
\includegraphics[width=\textwidth]{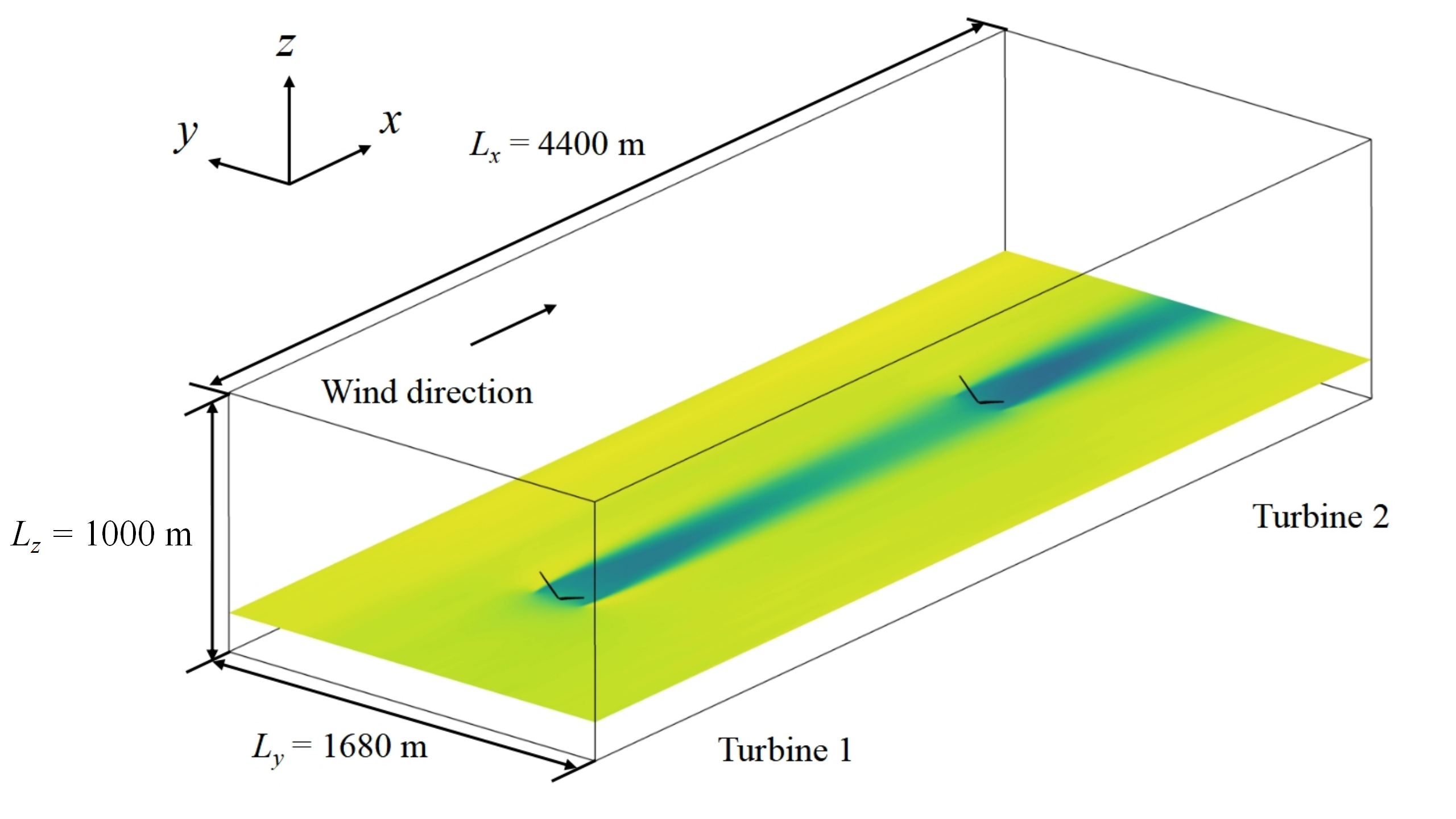}
\caption{Schematics of the computational domain of the two-turbine wind farm configuration. The simulation result of time-averaged wind speed on the hub-height plane is shown.}
\label{fig:computationalDomain}
\end{figure}

\subsubsection{Two-turbine wind farm configuration}

The first configuration is designed to simulate a two-wind-turbine-farm.
The simulations are conducted within a rectangular computational domain as depicted in Figure \ref{fig:computationalDomain}. For consistency with industry practices, subsequent spacing-related measurements are also expressed in terms of D, which denotes the rotor diameter of the turbine. The computational domain is defined with respect to the largest wind turbine employed in this work, \textit{i.e.}, the IEA 15 MW wind turbine with a rotor diameter ($D$) of 240 m. The extent of the computational domain is 4440 m ($18.5D$) in the streamwise direction, 1680 m ($7D$) in the transversal direction and 1000 m in the vertical direction. The streamwise distance between the inlet and the first wind turbine is 840 m ($3.5D$), leaving a downstream region of 3600 m ($15D$) for the wake. The turbine array is placed along the centerline of the domain. The grid spacing in the streamwise direction is uniform, with a spatial interval of $\Delta x = 12~\text{m}~(D/20)$. The grid in the transversal direction has a uniform interval of $\Delta y = 6 ~\text{m}~(D/40)$. In the vertical direction, the grid is uniform in the range of $ 0 \le z \le 480~\text{m} $ with $\Delta z = 6~\text{m}~(D/40)$ and is gradually stretched out until the top of the computational domain. The total number of grid is $N_x \times N_y \times N_z = 371 \times 281 \times 105 \approx $ 11 million. Previous work has demonstrated that such a spatial resolution is sufficient to obtain mesh-independent results for both first and second-order turbulence statistics in the far wake \citep{li2021large}. 

\subsubsection{Five-turbine wind farm configuration}

\begin{figure}[H]
\centering
\includegraphics[width=\textwidth]{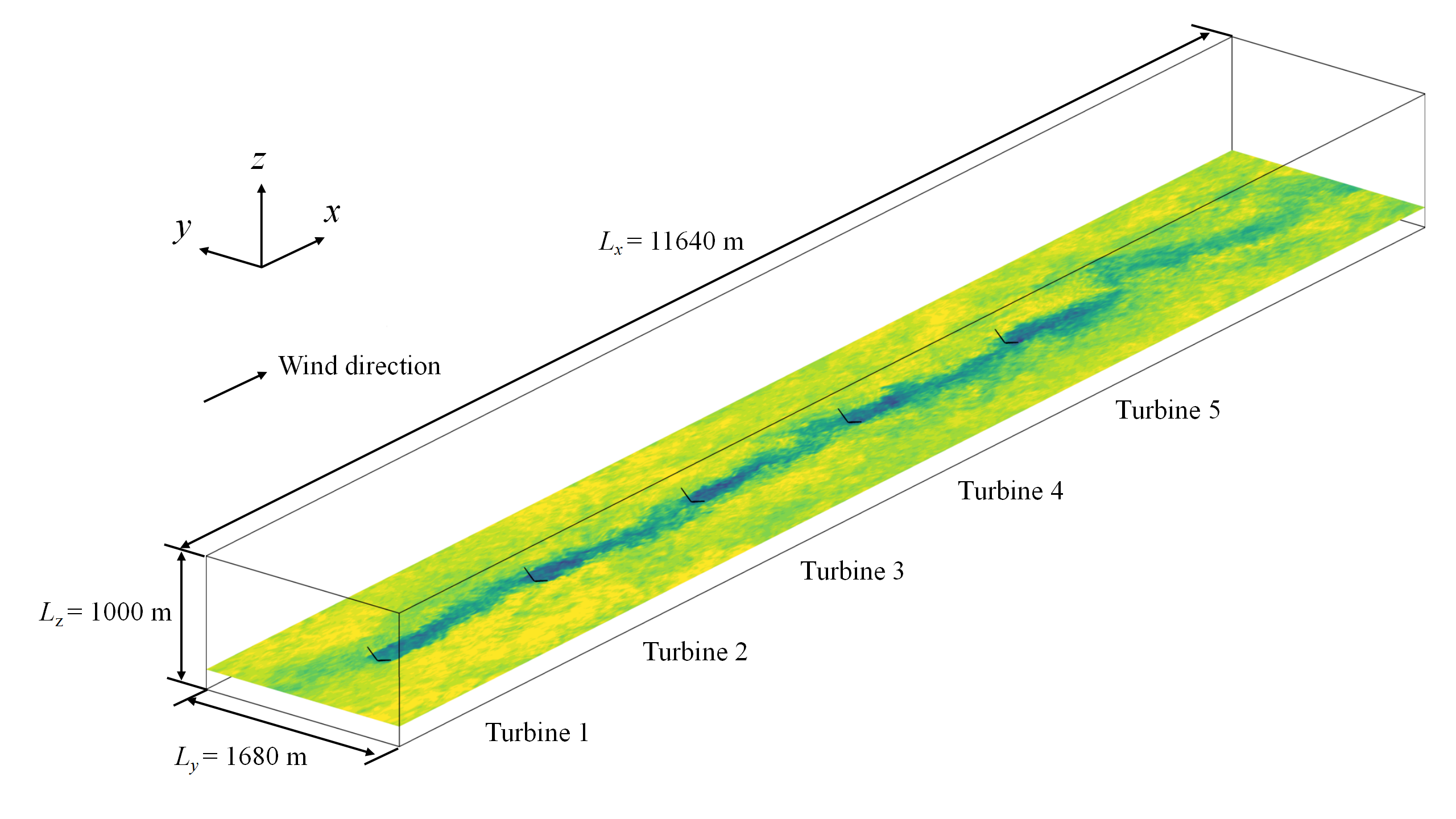}
\caption{Schematics of the computational domain of the five-turbine wind farm configuration, along with the simulation result of instantaneous wind speed on the hub-height plane.}
\label{fig:computationalDomain_revise}
\end{figure}

A wind farm of five DTU 10 MW turbines (Figure \ref{fig:computationalDomain_revise}) is also considered as an extension of the two-turbine case. The dimension of the computational domain is $L_x \times L_y \times L_z = 11640~ \text{m} \times 1680 ~\text{m} \times 1000 ~\text{m}$, i.e., the computational domain is extended in the streamwise direction compared to the two-turbine case. All turbines are aligned with the mean wind direction and spaced equally in the streamwise direction. To evaluate the model’s predictive performance, we use three spacings—890 m, 1335 m, and 1780 m—corresponding to 5 $D$, 7.5 $D$, and 10 $D$ rotor diameters. The turbine’s yaw angle $\gamma$ is varied collectively over $\{0^\circ, 5^\circ, 10^\circ, 20^\circ\}$, to test the model when the inflow and the wakes are misaligned. The grid resolution matches that of the two-turbine case, with a total of 
$N_x \times N_y \times N_z = 971 \times 281 \times 105 \approx 29\text{ million nodes}. $

\subsubsection{Boundary conditions}
\begin{figure}[H]
\centering
\includegraphics[width=0.8\textwidth]{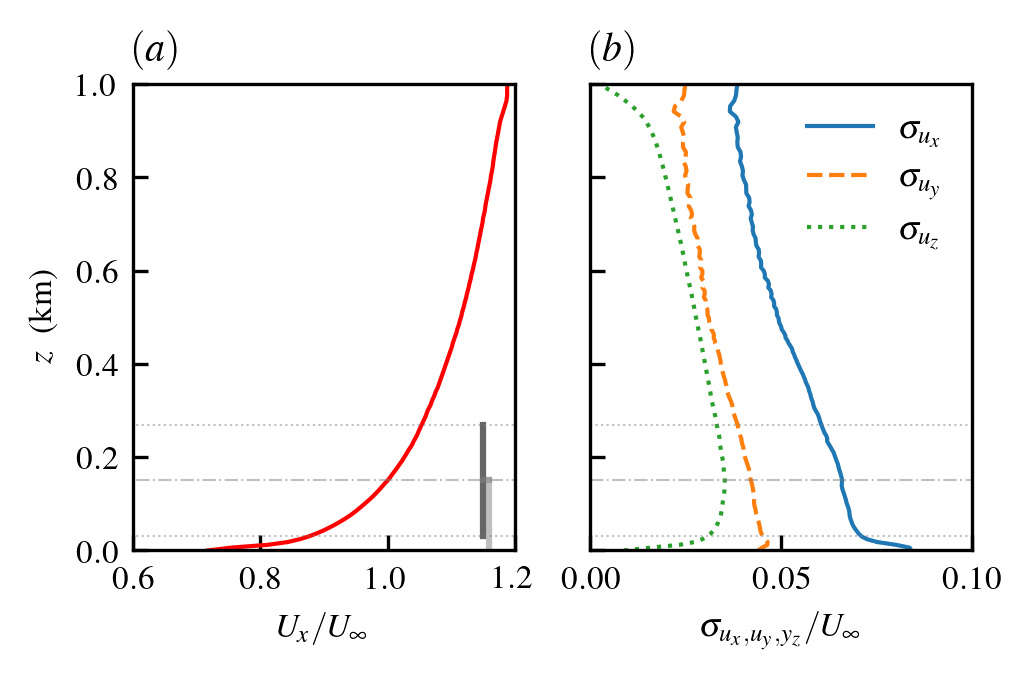}
\caption{Turbulent inflow statistics in the IEA 15 MW wind turbine case: the vertical profile of (a) the time-averaged streamwise velocity $U_x$ and (b) the standard deviation velocity components in three directions.}
\label{fig:inflow}
\end{figure}

The inlet boundary condition is imposed with a non-uniform and unsteady velocity field generated by a precursory LES which uses a half-channel configuration to mimic a fully developed neutral atmospheric boundary layer. The ground roughness length $z_0$ is employed to control the vertical velocity profile and the turbulence intensity in the precursory LES as well as the subsequent LES of wake flow. Figure \ref{fig:inflow} illustrates the time-averaged streamwise velocity ($U_x$) and the standard deviation of the velocity fluctuations at the inlet. At the turbine hub height ($z=150$ m), the turbulence intensities are measured as $I_x = 6.6\%$, $I_y = 4.2\%$, and $I_z = 3.5\%$, and the velocity is normalized by the free-stream velocity. Further details regarding the methodologies used for the precursor simulation and the interpolation techniques between the precursor and subsequent simulations are discussed extensively in our earlier publication \citep{li2021large}.

\subsection{Data Preprocessing}

To meet the requirements of this study, the LES results were time-averaged at each time step and subsequently interpolated onto a uniform grid. Furthermore, to extract a sufficient number of feature points using the SIFT method, the three-dimensional data were decomposed into a series of one-dimensional datasets along the turbine axial direction. Specifically, within the rotor diameter range, linear interpolation was performed based on the target turbine rotor diameter, while outside this range, data from the nearest grid points were used.

\section{Results of wind farm wake flow modeling}

The proposed method is applied to the wake flow fields collected in Section 3 to perform interpolation within the parameter space defined by turbine spacing, size, and wind direction misalignment. To evaluate its accuracy, six tests are designed. Test 1 serves as the baseline, using wake flow fields from two identical turbines with different spacings. Test 2 varies the turbine size, while Test 3 varies both turbine spacing and size. Test 4 involves the interpolation of wake flow fields for five turbines with different spacings. Test 5 also uses five turbines but with a small wind direction misalignment, and Test 6 examines large wind direction misalignment.

\subsection{Description of test cases}

The spacing of the turbines and the size of the turbines are two critical factors influencing the flow field of wind farms. Different turbine sizes correspond to varying wake intensities, while turbine spacings impact wake recovery.
\begin{figure}[H]
\centering
\includegraphics[width=0.6\textwidth]{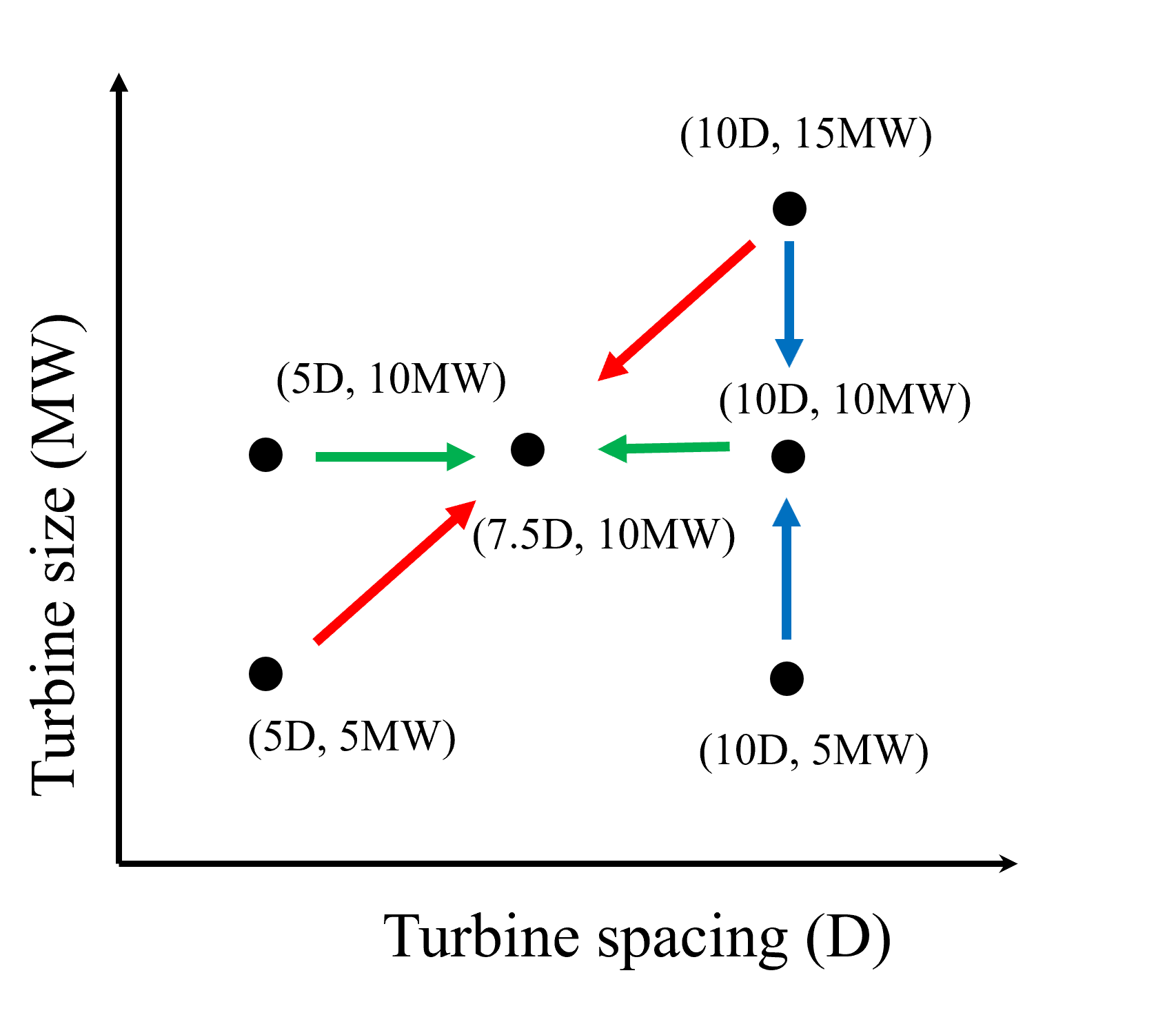}
\caption{A schematic diagram of test cases is shown. Green arrows represent Test 1, blue arrows represent Test 2, and red arrows represent Test 3. D represents the rotor diameter.}
\label{size-spacing space}
\end{figure}
In the parameter space defined by turbine spacing and size, the proposed data-driven modeling method is designed to interpolate new wake flow data from existing ones corresponding to different combinations of spacing and size. The functionality of this method is illustrated in Figure \ref{size-spacing space}, where arrows of different colors represent different tests. 

In addition to the aforementioned tests, the accuracy of the proposed modeling method was also validated in a larger wind farm comprising five turbines, with interpolations of turbine wake flow fields for different spacings and wind direction misalignments.

The performance of the proposed method is conducted through six tests, where all wake flow field data is generated using the method presented in Section 3.1. The specific parameters for those tests are detailed in Table \ref{table Tests}.

\begin{table}[H]
	\centering
	\caption{The specific parameters for the six experiments.}
	\label{table Tests}
	\renewcommand{\arraystretch}{1.5}
	\begin{tabular}{ccccc}
		\toprule
		\multirow{2}{*}{\makecell[c]{Tests}} & \multicolumn{2}{c}{Input} & \multicolumn{2}{c}{Target} \\
		\cline{2-5}
		& Turbine & Spacing/Degree & Turbine & Spacing/Degree \\
		\midrule
		
		\multirow{2}{*}{\makecell[c]{Test 1}} & 2 DTU 10MW & 890m/{$0^\circ$} & \multirow{2}{*}{2 DTU 10MW} & \multirow{2}{*}{1335m/{$0^\circ$}} \\
		\cline{2-3}
		& 2 DTU 10MW & 1780m/{$0^\circ$} & & \\
		\midrule
		
		\multirow{2}{*}{\makecell[c]{Test 2}} & 2 NREL 5MW & 1780m/{$0^\circ$} & \multirow{2}{*}{2 DTU 10MW} & \multirow{2}{*}{1780m/{$0^\circ$}} \\
		\cline{2-3}
		& 2 IEA 15MW & 1780m/{$0^\circ$} & & \\
		\midrule
		
		\multirow{2}{*}{\makecell[c]{Test 3}} & 2 NREL 5MW & 890m/{$0^\circ$} & \multirow{2}{*}{2 DTU 10MW} & \multirow{2}{*}{1335m/{$0^\circ$}} \\
		\cline{2-3}
		& 2 IEA 15MW & 1780m/{$0^\circ$} & & \\
		\midrule
		
		\multirow{2}{*}{\makecell[c]{Test 4}} & 5 DTU 10MW & 890m/{$0^\circ$} & \multirow{2}{*}{5 DTU 10MW} & \multirow{2}{*}{1335m/{$0^\circ$}} \\
		\cline{2-3}
		& 5 DTU 10MW & 1780m/{$0^\circ$} & & \\
		\midrule
		
		\multirow{2}{*}{\makecell[c]{Test 5}} & 5 DTU 10MW & 1780m/$0^\circ$ & \multirow{2}{*}{5 DTU 10MW} & \multirow{2}{*}{1780m/$5^\circ$} \\
		\cline{2-3}
		& 5 DTU 10MW & 1780m/$10^\circ$ & & \\
		\midrule
		
		\multirow{2}{*}{\makecell[c]{Test 6}} & 5 DTU 10MW & 1780m/$0^\circ$ & \multirow{2}{*}{5 DTU 10MW} & \multirow{2}{*}{1780m/$10^\circ$} \\
		\cline{2-3}
		& 5 DTU 10MW & 1780m/$20^\circ$ & & \\
		\bottomrule
	\end{tabular}
\end{table}

Altering spacing is a fundamental requirement in wind farm design. The proposed modeling method is tested under flow field input conditions with different spacings to verify its accuracy through Test 1. Wake flow data from two aligned DTU 10MW turbines, spaced at 890 m ($5D$) and 1780 m ($10D$) respectively, are utilized as input data. Meanwhile, the wake flow field associated with a spacing of 1335 m ($7.5D$) is designated as the target. 

In real wind farm design, different turbine sizes are typically used. In Test 2, wake field data from two different turbine sizes are utilized to evaluate the accuracy of the proposed modeling method. Wake flow data from two aligned NREL 5MW turbines and two aligned IEA 15MW turbines with a spacing of 1780 m ($10D$) are used as input. Meanwhile, the wake flow field from two aligned DTU 10MW turbines with a spacing of 1780 m ($10D$) serves as the target.

Simultaneously altering turbine types and spacing presents a more intricate challenge for wind farm optimization. Test 3 is designed to validate the applicability of the proposed method under these conditions. Wake flow fields from two aligned NREL 5MW turbines with a spacing of 890 m ($5D$) and from two aligned IEA 15MW turbines with a spacing of 1780 m ($10D$) are used as inputs. Meanwhile, the wake flow field from two aligned DTU 10MW turbines with a spacing of 1335 m ($7.5D$) is utilized as the target.

Wake superposition is a non-negligible physical phenomenon in large wind farms. The wakes generated by upstream turbines interact and accumulate downstream, resulting in more complex wake flow structures that are challenging to predict accurately. In Test 4, the proposed modeling method was evaluated for its interpolation accuracy in predicting wake fields with multiple overlapping turbine wakes. By varying the turbine spacing, the degree to which downstream turbines were affected by the wakes of upstream turbines was controlled. In this case, five DTU 10 MW turbines were used to generate the wake flow fields. Wake flow fields with turbine spacings of 890 m and 1780 m were used as inputs, while the wake flow field with a spacing of 1335 m served as the target. This case demonstrates the potential of the proposed method for application in large-scale wind farms.

Wind direction misalignment has a significant impact on turbine wake behaviors by altering wake deflection, overlap and recovery, thereby affecting the performance of downstream turbines and the overall power output of a wind farm. In Tests 5 and 6, the applicability of the proposed modeling method under varying wind direction misalignment was evaluated. Both tests employed five DTU 10 MW turbines with a uniform spacing of 1780 m. Test 5 assessed the method’s performance under small wind direction variations, using wake flow fields corresponding to wind direction misalignment of $0^\circ$ and $10^\circ$ as inputs, with the wake flow field at $5^\circ$ serving as the target. In contrast, Test 6 evaluated the method under large wind direction differences, using wake flow fields at $0^\circ$ and $20^\circ$ as inputs and the wake flow field at $10^\circ$ as the target. The results of these tests demonstrate the robustness and applicability of the proposed method in capturing wake variations under different wind direction misalignment scenarios.

The targets for all six tests correspond to the DTU 10MW turbine, allowing for a comparison of results across the different tests.

\subsection{Test 1: different spacings}

\begin{figure}[H]
\centering
\includegraphics[width=1\textwidth]{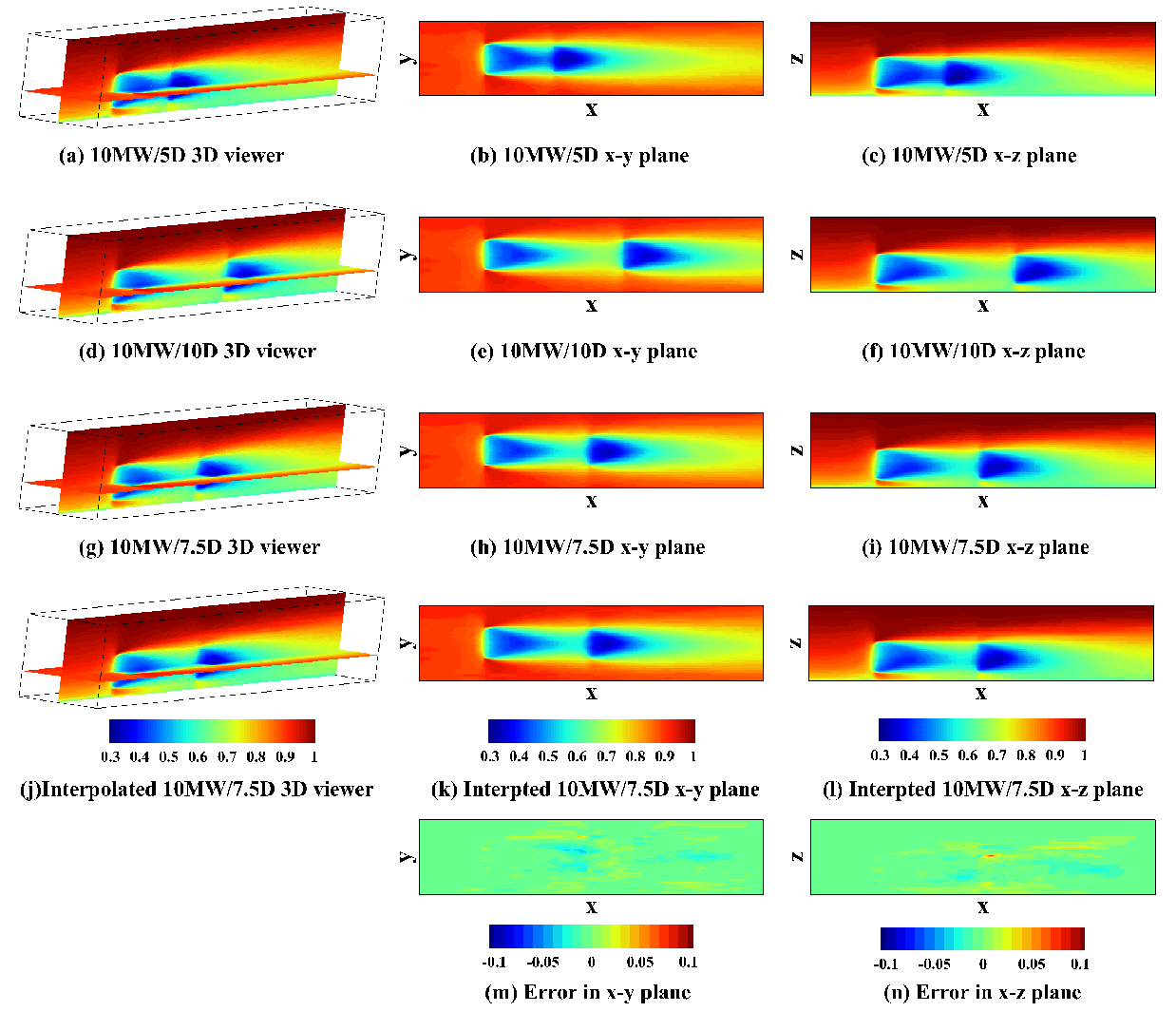}
\caption{Interpolation results of Test 1.}
\label{exp 1}
\end{figure}

Firstly, interpolation of flow fields with varying spacings is tested, indicating a good agreement between the interpolated (Figure \ref{exp 1}j, \ref{exp 1}k, \ref{exp 1}l) and the target (Figure \ref{exp 1}g, \ref{exp 1}h, \ref{exp 1}i) wake flow field, both in the x-y plane and the x-z plane. The error maps shown in Figure \ref{exp 1}m and \ref{exp 1}n highlight that significant discrepancies primarily occur at near-wake and turbine blade tips. In these regions, the flow field experiences abrupt changes, which present challenges in accurately identifying feature points and consequently introduce bias.

Between the turbines and the ground, the blockage effect induces a notable increase in velocity, which can be successfully interpolated by the present scheme, demonstrating its capacity to capture flow field characteristics.

\begin{figure}[H]
\centering
\includegraphics[width=0.85\textwidth]{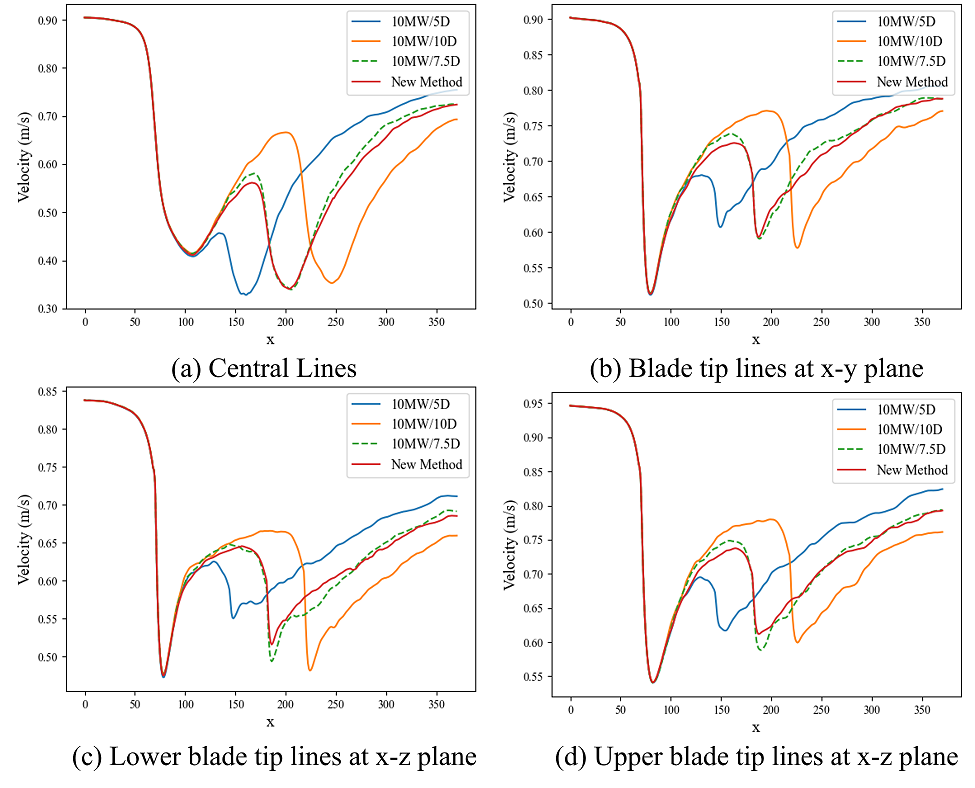}
\caption{Interpolated streamwise velocity of Test 1 along specific streamlines.}
\label{test1_lines}
\end{figure}

Streamline comparisons at specific locations within the flow field are also conducted, as presented in Figure \ref{test1_lines}. It can be observed that the interpolated results match well with the target in terms of both wake decay and wake recovery.

\subsection{Test 2: different sizes}

\begin{figure}[H]
\centering
\includegraphics[width=1\textwidth]{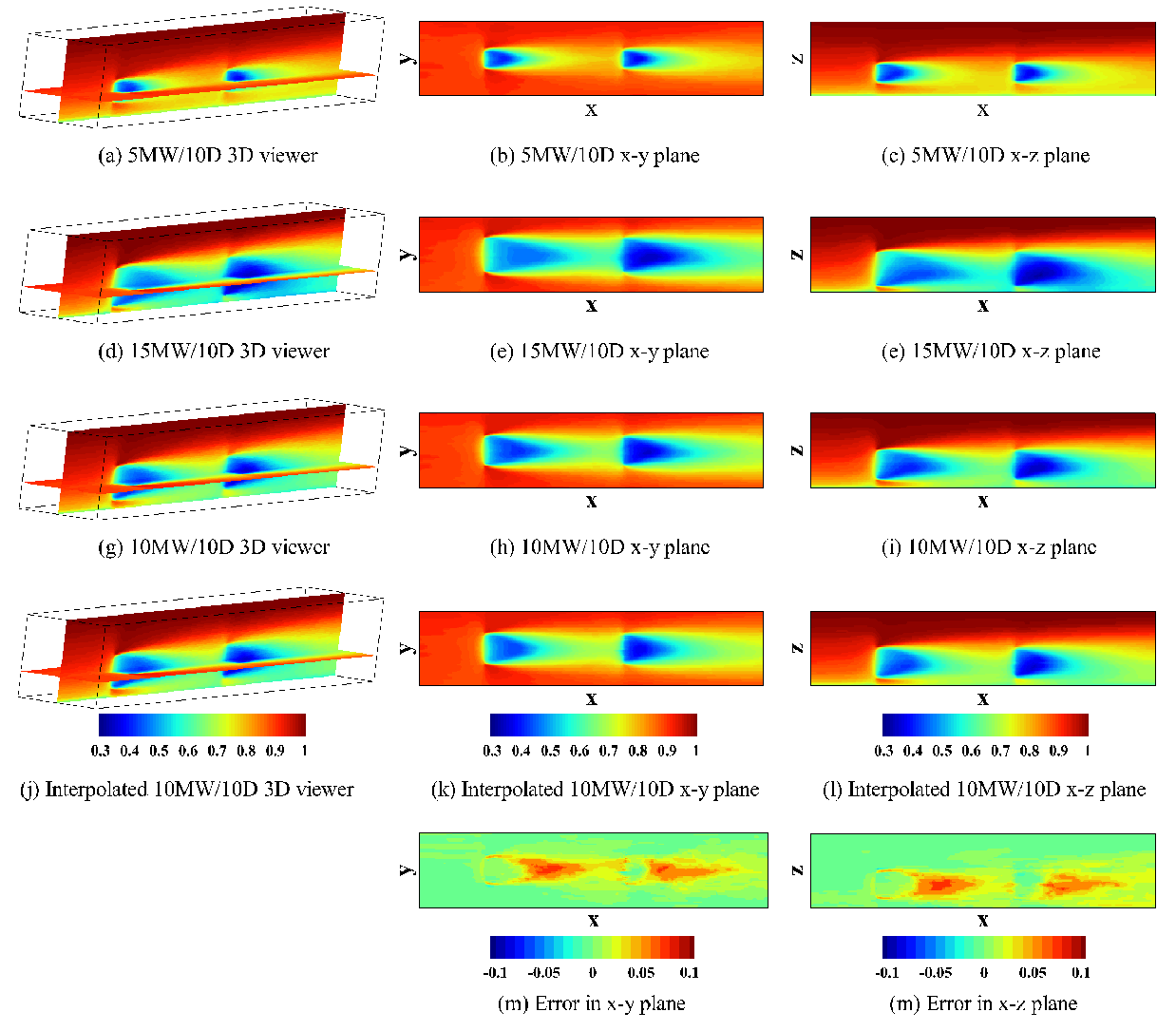}
\caption{Interpolation results of Test 2.}
\label{exp 2}
\end{figure}

\begin{figure}[H]
\centering
\includegraphics[width=0.85\textwidth]{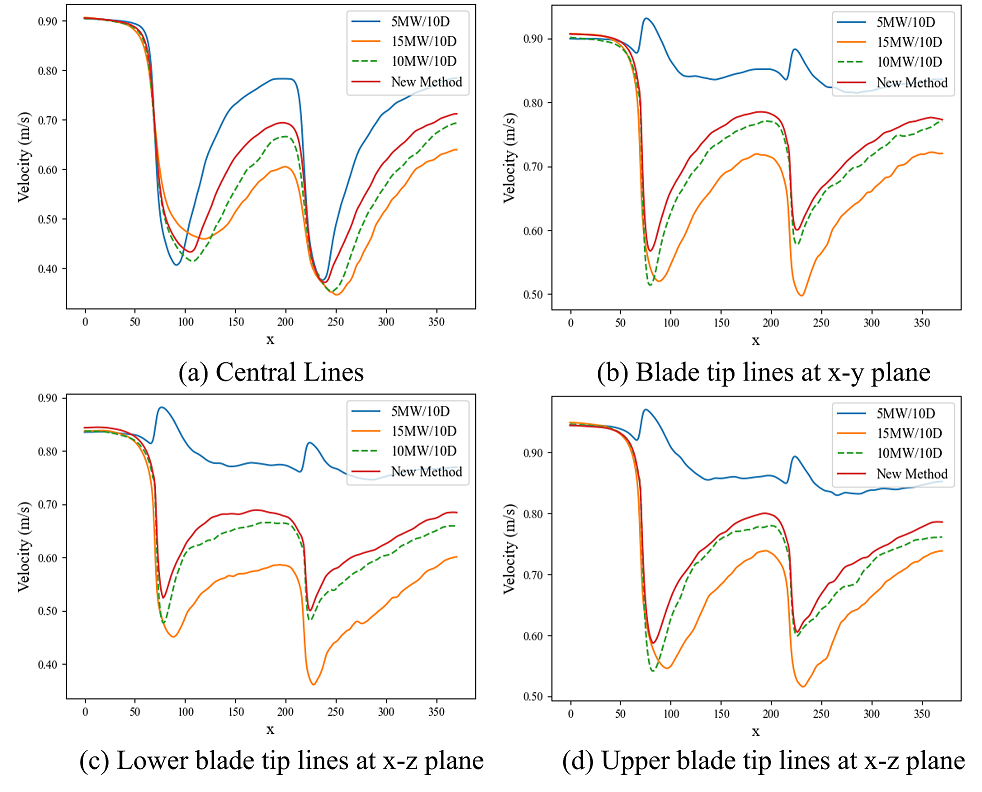}
\caption{Interpolation streamwise velocity of Test 2 along specific streamlines.}
\label{test2_lines}
\end{figure}

In wind farms, turbines with various sizes exhibit distinct fluid dynamics such as turbulence, wake intensity, wake influence range, and wake recovery. LES flow fields of the NREL 5MW and IEA 15MW turbines are used as inputs, and the interpolation results are validated against the DTU 10MW LES data. As shown in Figure \ref{exp 2}, significant differences in wake fields arise due to variations in turbine size. The proposed method effectively interpolates the shape of the target wake field. Figure \ref{exp 2}m and \ref{exp 2}n indicate that normalized velocity errors are generally within 0.05, demonstrating high accuracy, with errors primarily occurring in the far wake and increasing with greater distances. This is likely due to the nonlinear relationship between turbine size and the wake flow field, which cannot be accurately modeled by linear functions. A similar phenomenon is reflected in the central line in Figure \ref{test2_lines}, where the streamline for the 10 MW turbine is closer to the 15MW streamline rather than being positioned at the average location between the 5 MW and 15 MW streamlines.

Compared to Test 1, Test 2 exhibits significantly larger errors, indicating that changes in turbine size introduce greater complexity to the flow field variations.

\subsection{Test 3: different sizes and spacings}

\begin{figure}[H]
\centering
\includegraphics[width=1\textwidth]{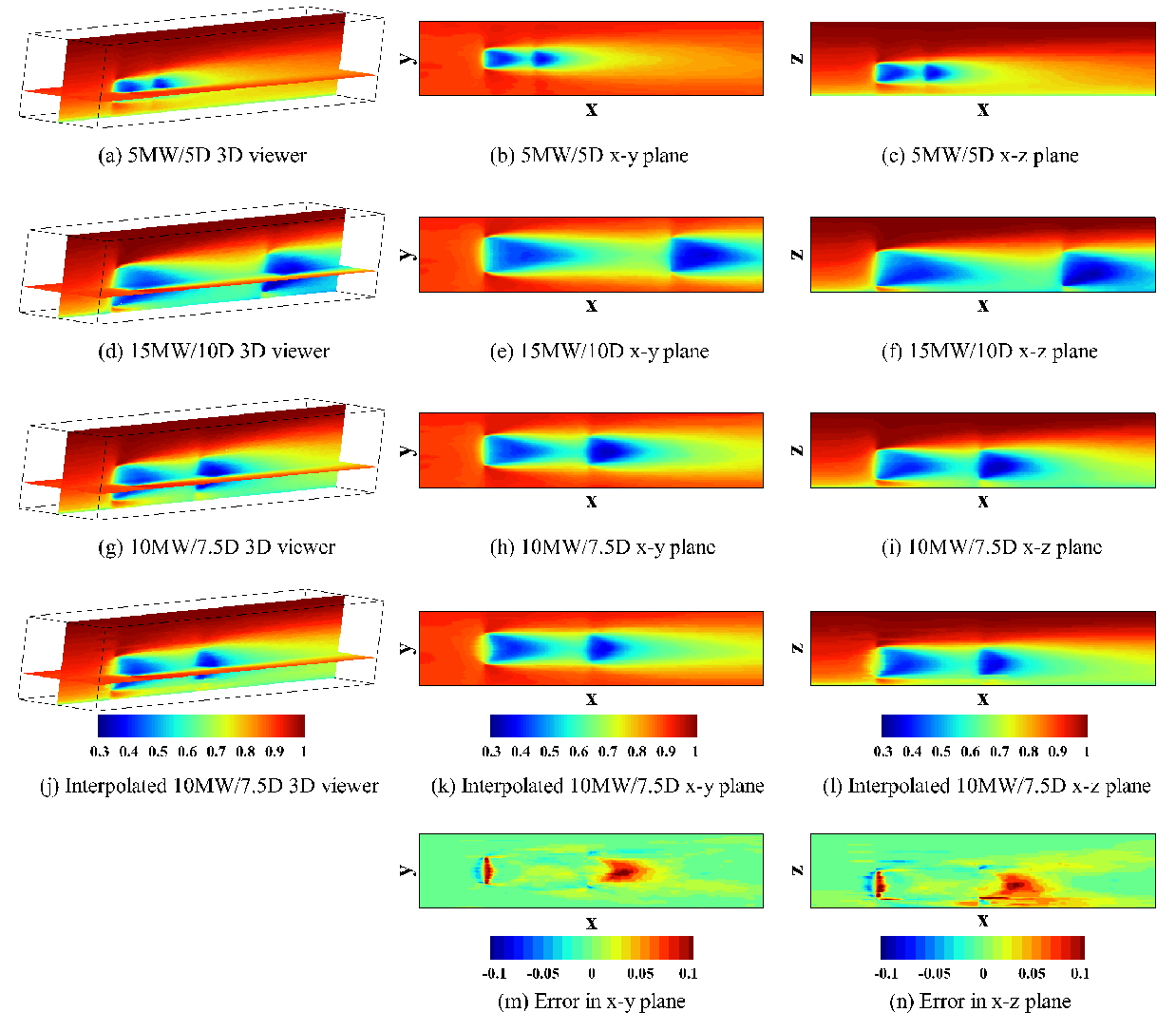}
\caption{Interpolation results of Test 3.}
\label{exp 3}
\end{figure}

In real wind farm designs, different sizes are often associated with different spacing. Such a simultaneous variation of size and spacing significantly increases modeling complexity due to the challenges in feature identification and matching. Nevertheless, Figure \ref{exp 3} demonstrates good consistency between the interpolated result and the target. The proposed algorithm can accurately interpolate turbine positions, velocity field extremes, and wake shapes within the flow field. However, as depicted in Figure \ref{exp 3}m and \ref{exp 3}n, deviations remain noticeable around the blade and the downstream area.

\begin{figure}[H]
\centering
\includegraphics[width=0.85\textwidth]{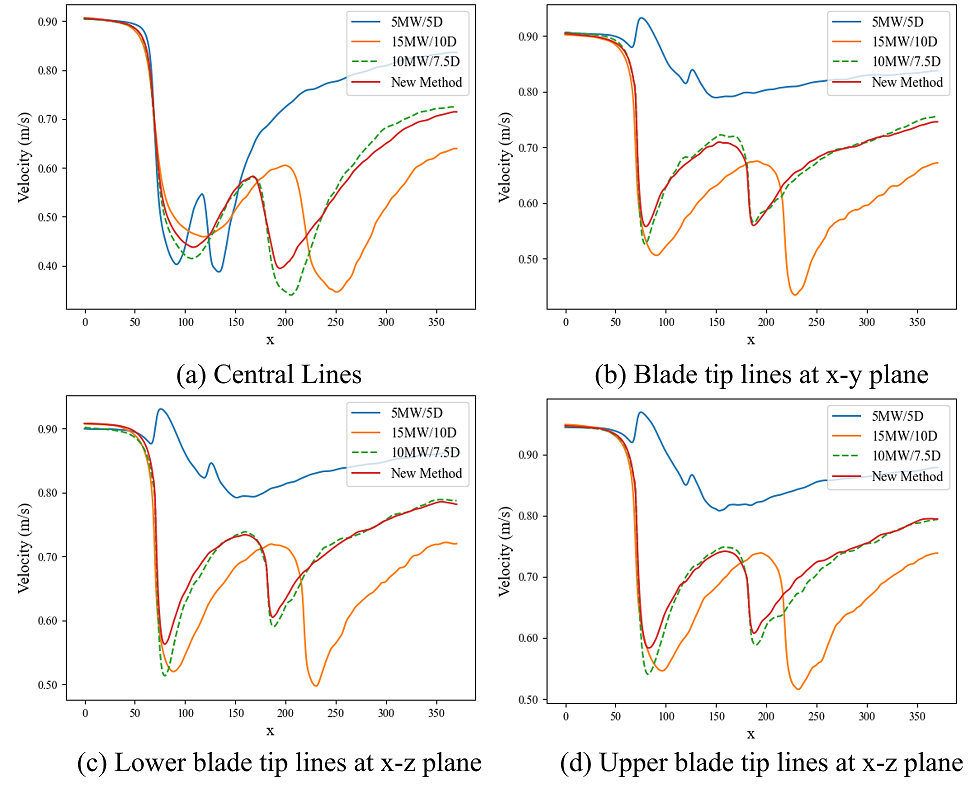}
\caption{Interpolation streamwise velocity of Test 3 along specific streamlines.}
\label{test3_lines}
\end{figure}

Detailed comparisons of streamline results are also presented in Figure \ref{test3_lines}. Errors increase significantly near the second turbine, which may be attributed to the simultaneous changes in both turbine size and spacing. This added complexity complicates the determination and matching of feature points, thereby making it more difficult to interpolate the position of the second turbine. Additionally, the nonlinear relationship between turbine size and the wake field results in larger errors at greater distances from the turbine.

\subsection{Test 4: different spacings with five turbines}

\begin{figure}[H]
\centering
\includegraphics[width=1\textwidth]{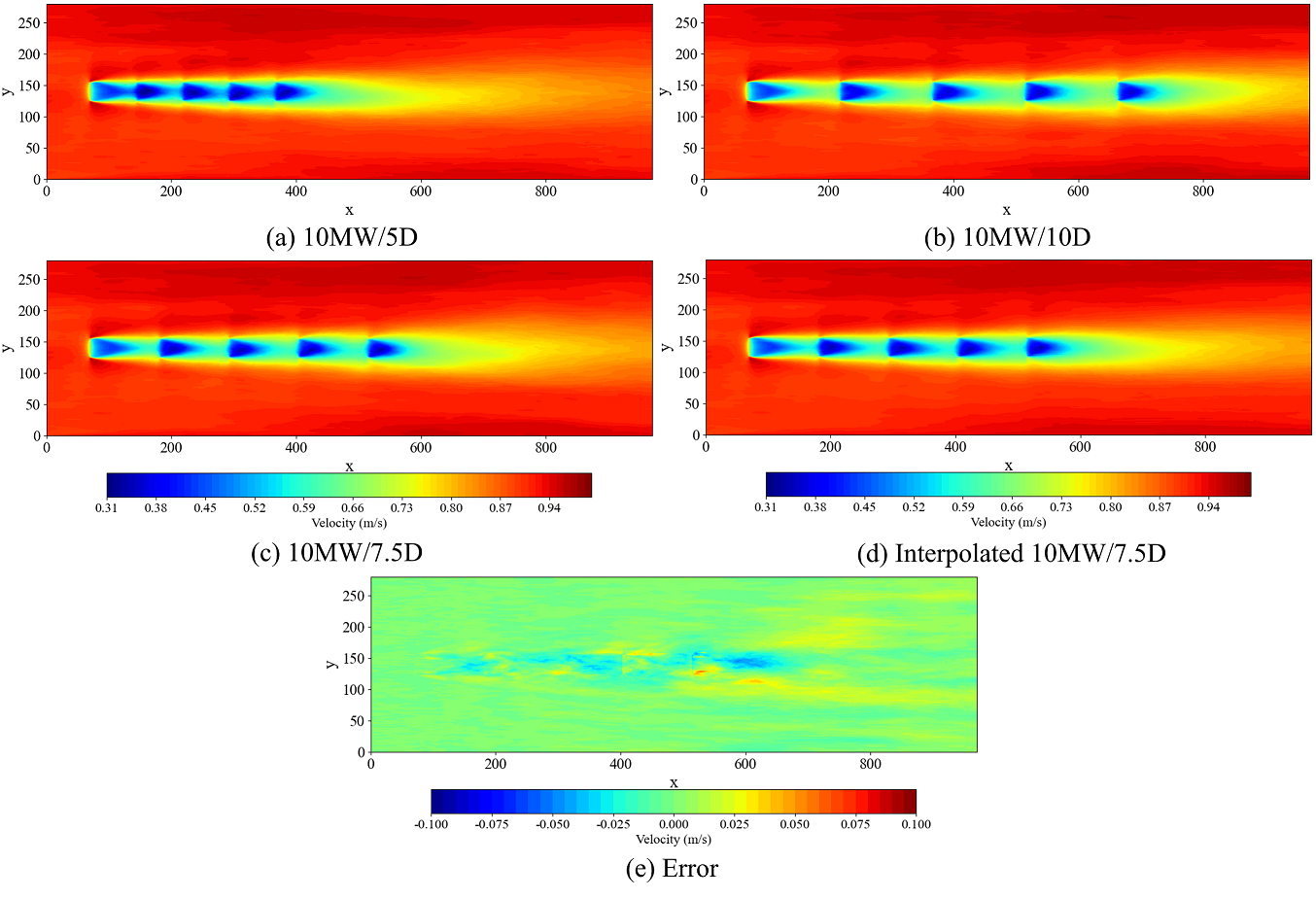}
\caption{Interpolation results of Test 4 at x-y plane.}
\label{exp 4 xy}
\end{figure}

\begin{figure}[H]
\centering
\includegraphics[width=1\textwidth]{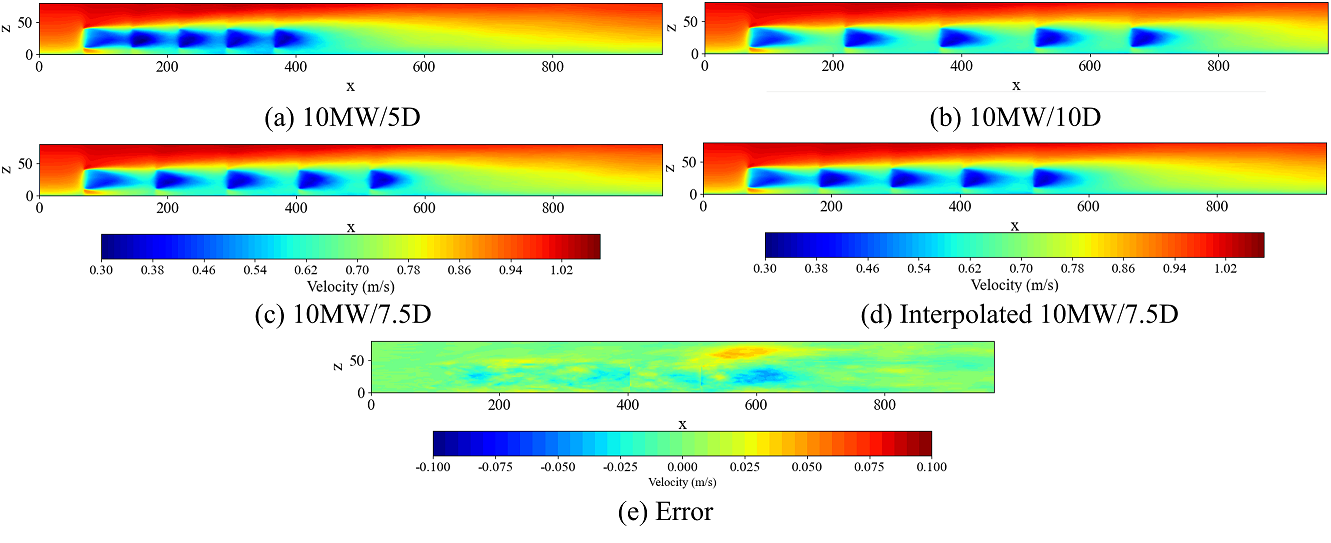}
\caption{Interpolation results of Test 4 at x-z plane.}
\label{exp 4 xz}
\end{figure}

To evaluate the applicability of the proposed method for wake modeling in large wind farms, its interpolation performance was tested on wake flow fields generated by five turbines aligned in a single row. Specifically, Figures \ref{exp 4 xy} and \ref{exp 4 xz} demonstrate that the interpolated results closely match the target wake flow fields in both the x-y plane and x-z plane. The method accurately captures key wake features, including overlap, lateral expansion, vertical structure, and downstream recovery, indicating its effectiveness in reconstructing complex wake superposition patterns under varying turbine spacings. Additionally, the error distributions in both planes are relatively small and evenly distributed, suggesting that the interpolation introduces minimal localized deviations and confirming the model’s capability to predict wake fields across different spatial configurations.

\begin{figure}[H]
\centering
\includegraphics[width=0.85\textwidth]{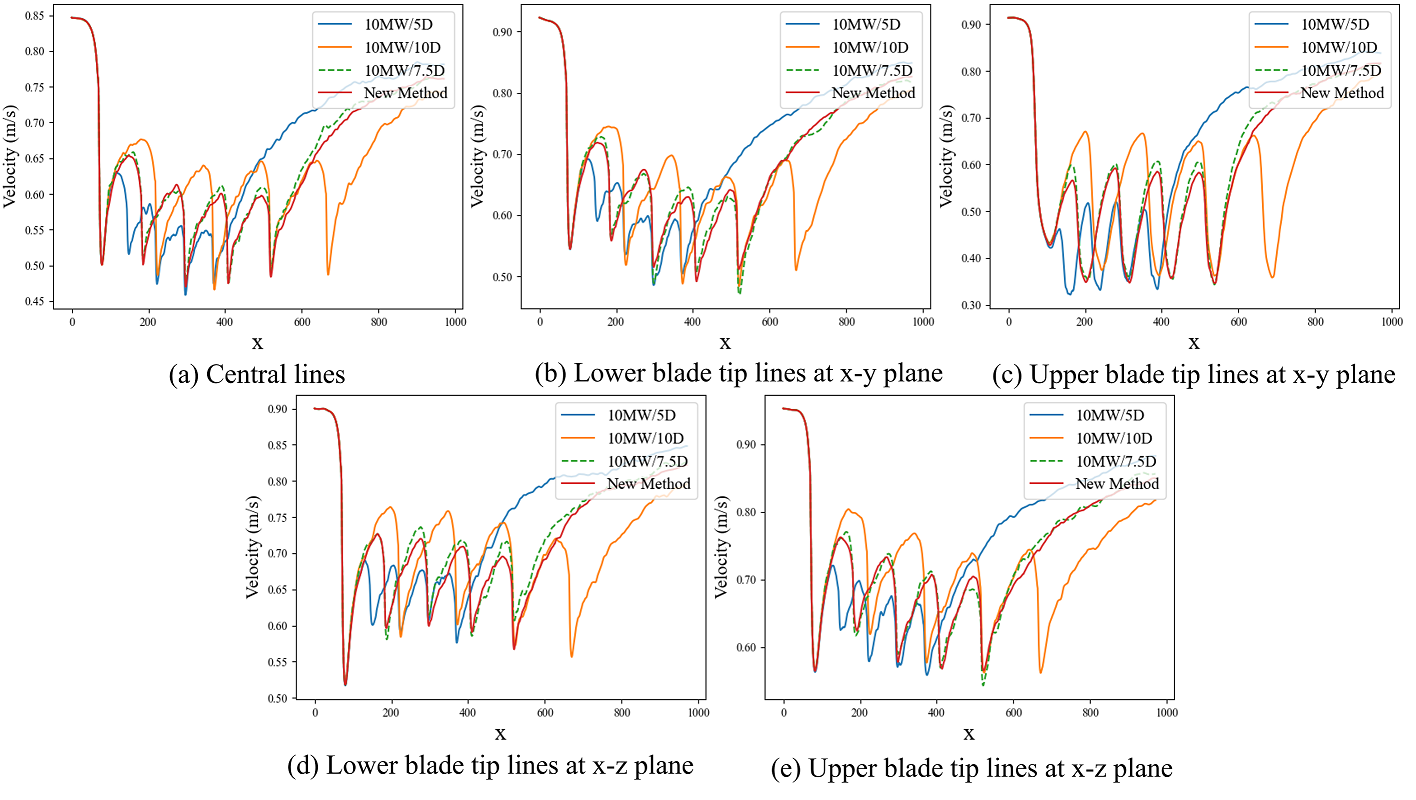}
\caption{Interpolation streamwise velocity of Test 4 along specific streamlines.}
\label{exp4_lines}
\end{figure}

Figure \ref{exp4_lines} shows the comparison of velocity profiles along different streamwise lines for turbine spacings of $5D$, $7.5D$, and $10D$, where $7.5D$ serves as the interpolation target. Across all positions, including the turbine centerline, blade tips on the x-y plane, and blade tips on the x-z plane, the interpolated results (red lines) closely match the actual $7.5D$ spacing data (green dashed lines). This demonstrates that the proposed method can accurately reconstruct key wake characteristics, such as velocity deficits and recovery trends, at various spatial locations, highlighting its robustness and applicability for wake field modeling under different turbine spacing conditions in large wind farms.

\subsection{Test 5: small wind direction misalignment}

\begin{figure}[H]
\centering
\includegraphics[width=1\textwidth]{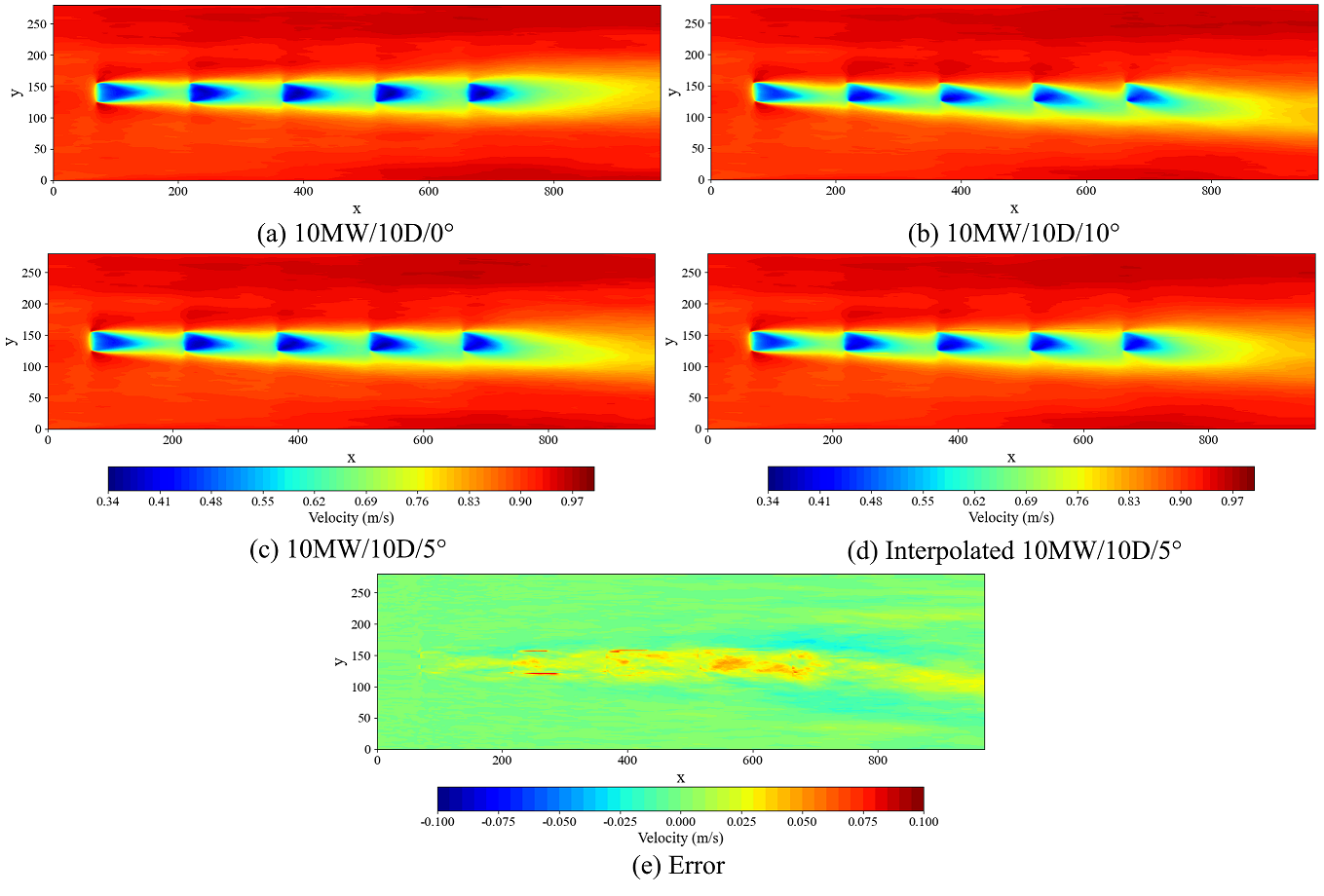}
\caption{Interpolation results of Test 5 at x-y plane.}
\label{exp 5 xy}
\end{figure}

\begin{figure}[H]
\centering
\includegraphics[width=1\textwidth]{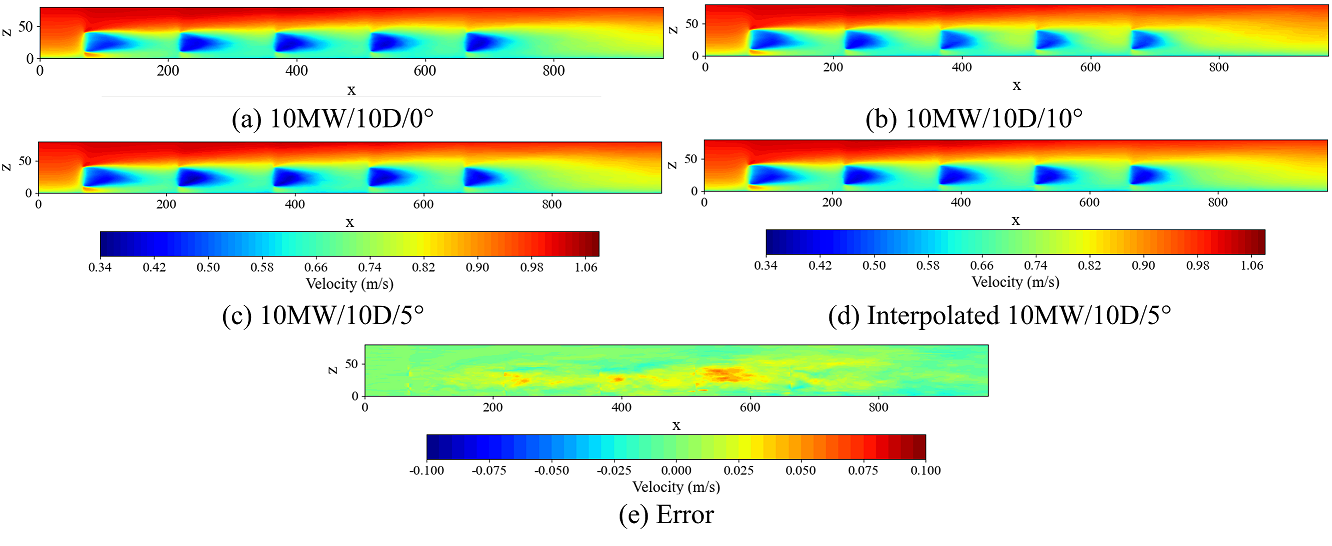}
\caption{Interpolation results of Test 5 at x-z plane.}
\label{exp 5 xz}
\end{figure}

\begin{figure}[H]
\centering
\includegraphics[width=0.85\textwidth]{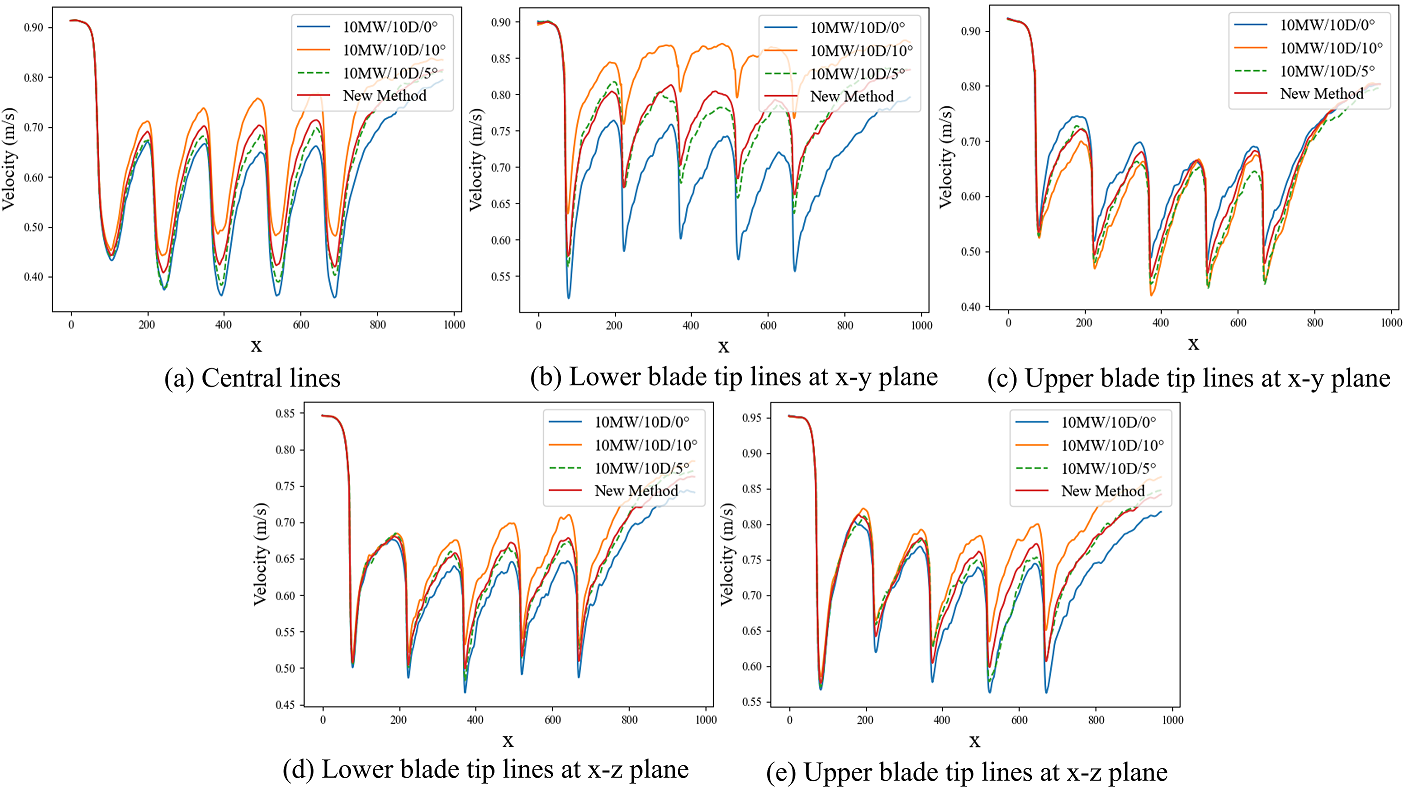}
\caption{Interpolation streamwise velocity of Test 5 along specific streamlines.}
\label{exp5_lines}
\end{figure}

For wake flow fields involving multiple turbines, we evaluated not only the interpolation accuracy of the proposed method under different turbine spacings but also its performance under various degrees of wind direction misalignment. In this study, both small- and large-angle wind direction misalignment scenarios were examined. In the small-angle case, wake flow fields at wind directions of $0^\circ$ and $10^\circ$ were used as inputs to interpolate the wake field at $5^\circ$ for the five-turbine configuration. In the large-angle case, wake fields at $0^\circ$ and $20^\circ$ served as inputs to interpolate the wake field at $10^\circ$. Figures \ref{exp 5 xy} and \ref{exp 5 xz} present the interpolation results on the x–y and x–z planes, respectively, for the small-angle scenario, while Figure \ref{exp5_lines} compares the velocity profiles at specific locations. These results demonstrate that the proposed method achieves high interpolation accuracy under small wind direction misalignment. It can be observed that when the degree of wind direction misalignment is less than $10^\circ$, the wake deflection remains small, and the wake flow fields exhibit high similarity, which ensures the accuracy of the proposed method.

\subsection{Test 6: large wind direction misalignment}

\begin{figure}[H]
\centering
\includegraphics[width=1\textwidth]{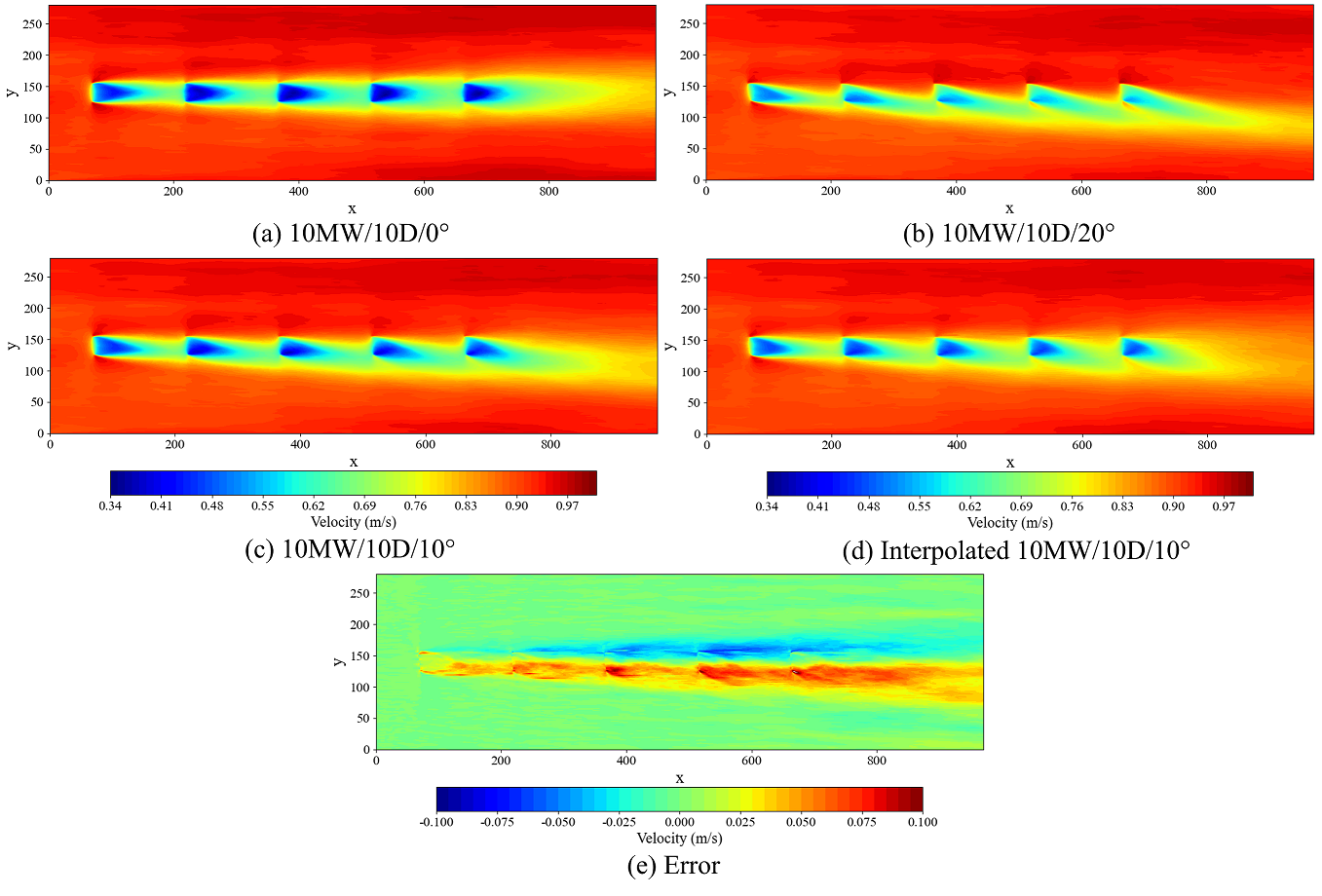}
\caption{Interpolation results of Test 6 at x-y plane.}
\label{exp 6 xy}
\end{figure}

\begin{figure}[H]
\centering
\includegraphics[width=1\textwidth]{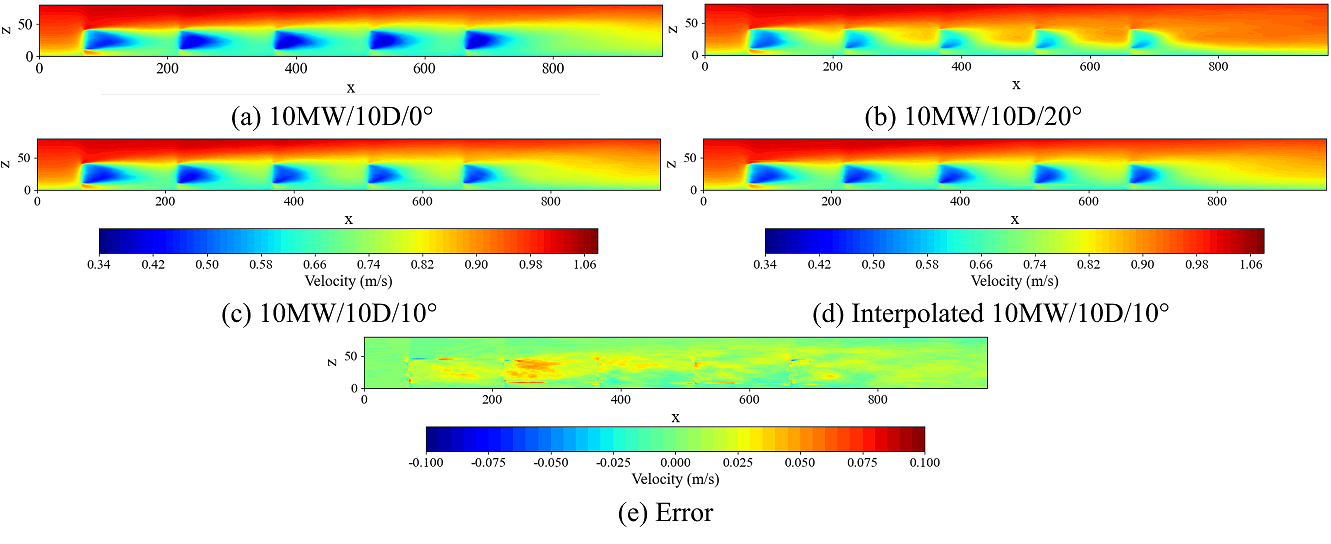}
\caption{Interpolation results of Test 6 at x-z plane.}
\label{exp 6 xz}
\end{figure}

\begin{figure}[H]
\centering
\includegraphics[width=0.85\textwidth]{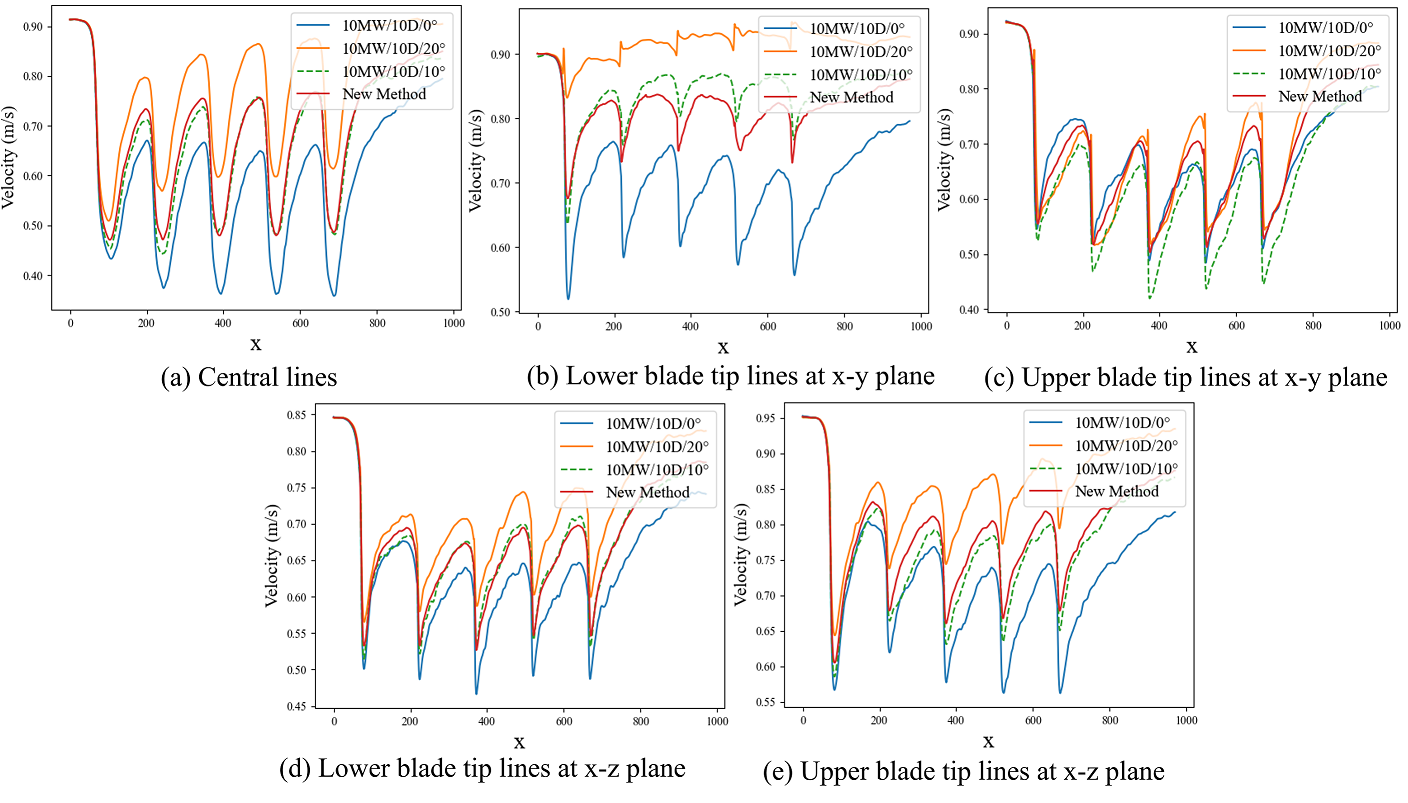}
\caption{Interpolation streamwise velocity of Test 6 along specific streamlines.}
\label{exp6_lines}
\end{figure}

Figures \ref{exp 6 xy}, \ref{exp 6 xz}, and \ref{exp6_lines} indicate that when the wind direction misalignment increases to $20^\circ$, the wake structure undergoes significant changes compared to the $0^\circ$ case. This is particularly evident in the x–y plane, where noticeable variations occur in the wake deficit and recovery patterns. Distinct velocity peaks and troughs appear near the lower blade tips of the downstream turbines, and the extent of these peak regions further expands downstream. Consequently, the similarity between the wake structures of the downstream turbines and those of the first-row turbines, as well as wakes under small wind direction misalignments, is greatly reduced, leading to decreased interpolation accuracy. Nevertheless, the proposed method still achieves high accuracy under these challenging conditions.

\subsection{Interpolation Accuracy and Deviation Analysis}

\begin{figure}[H]
\centering
    \includegraphics[width=0.85\textwidth]{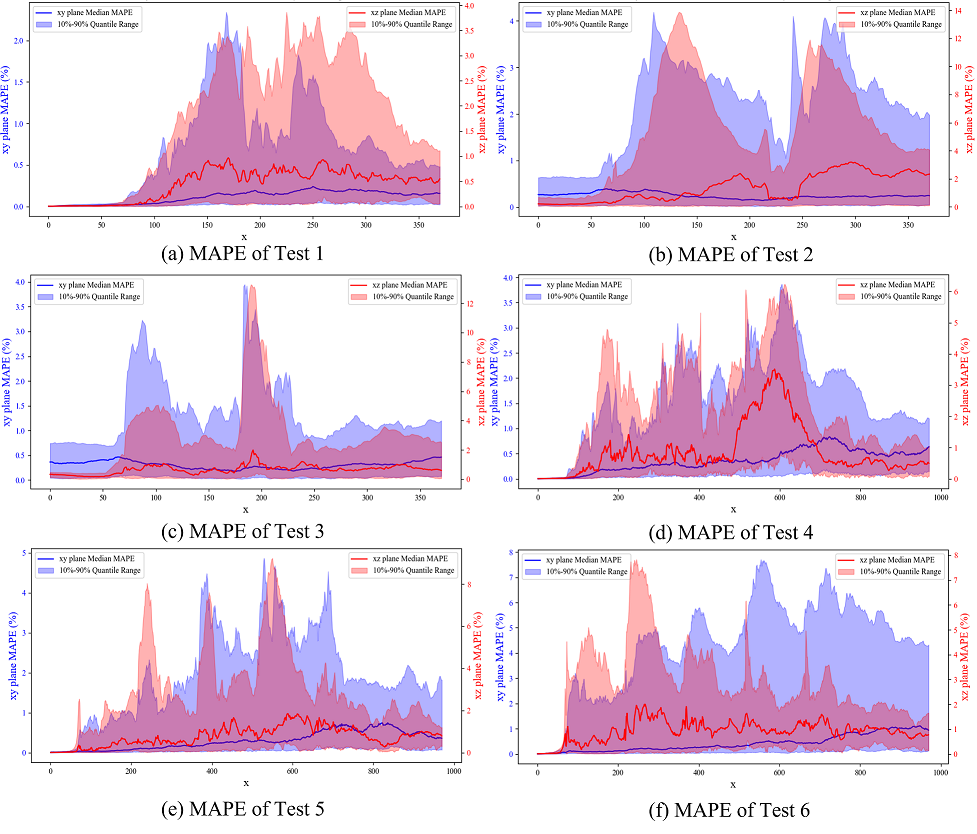}
\caption{MAPE distribution along the x-direction in x-y and x-z planes. Solid lines show median values, and shaded areas indicate the 10–90\% quantile ranges, reflecting interpolation variability in each plane.}
\label{quantile_range}
\end{figure}

Table \ref{table_error} presents the absolute and relative errors of six tests across different planes, while Figure \ref{quantile_range} illustrates the distribution of MAPE along the x-direction. For the two-turbine cases, the interpolation error in the x-z plane exceeds that in the x-y plane by more than a factor of two. This can be primarily attributed to the greater variation in wake structures among different tests in the x-z plane compared to the x-y plane, resulting in reduced similarity and thus higher interpolation errors. Although the errors show a gradual increase from Test 1 to Test 3, the overall error remains below 2.28\%, indicating satisfactory interpolation accuracy.

\begin{table}[H]
	\centering
	\caption{MAE and MAPE of modeling results}
	\label{table_error}
	\renewcommand{\arraystretch}{1.5}
	\begin{tabular}{ccccc}
		\toprule
		Tests & Planes & \makecell[c]{MAE} & \makecell[c]{MAPE ($\%$)} \\
		\midrule
		
		\multirow{2}{*}{\makecell[c]{Test 1}} & x-y & 0.0022 & 0.29 \\
		\cline{2-4}
		& x-z & 0.0045 & 0.68 \\
		\midrule
		
		\multirow{2}{*}{\makecell[c]{Test 2}} & x-y & 0.0052 & 0.77 \\
		\cline{2-4}
		& x-z & 0.0132 & 2.18 \\
		\midrule
		
		\multirow{2}{*}{\makecell[c]{Test 3}} & x-y & 0.0046 & 0.65 \\
		\cline{2-4}
		& x-z & 0.0143 & 2.28 \\
		\midrule
		
		\multirow{2}{*}{\makecell[c]{Test 4}} & x-y & 0.0049 & 0.63 \\
		\cline{2-4}
		& x-z & 0.0076 & 1.11 \\
		\midrule
		
		\multirow{2}{*}{\makecell[c]{Test 5}} & x-y & 0.0049 & 0.68 \\
		\cline{2-4}
		& x-z & 0.0078 & 1.19 \\
		\midrule
		
		\multirow{2}{*}{\makecell[c]{Test 6}} & x-y & 0.0105 & 1.41 \\
		\cline{2-4}
		& x-z & 0.0092 & 1.30 \\
		\bottomrule
	\end{tabular}
\end{table}

In the five-turbine cases, the errors for Test 4 through Test 6 are all below 2\%, demonstrating the robustness and applicability of the proposed method in complex multi-turbine wake fields. A comparison between Test 1 and Test 4 indicates that while the overall error slightly increases with the number of turbines, the magnitude of this increase is minimal. Furthermore, a comparison between Test 5 and Test 6 reveals that under small wind direction misalignment, the interpolation error in the x-z plane is approximately 1.75 times greater than that in the x-y plane. However, when the wind direction misalignment increases to $20^\circ$, the interpolation error in the x-y plane rises sharply, surpassing that in the x-z plane. This is primarily due to the significant changes in wake structures in the x-y plane, which substantially reduce the horizontal similarity between the $0^\circ$ and $20^\circ$ wake fields.

The distribution of MAPE along the x-direction reveals that interpolation errors are primarily concentrated in regions exhibiting the highest wake dissipation. Furthermore, the error range in the x-z plane is notably larger than that in the x-y plane, with Tests 2 and 3 showing the greatest errors, followed by Tests 5 and 6. These results suggest that variations in turbine size and wind direction can significantly impact interpolation accuracy. This may be attributed to the reduced similarity between the two input wake fields in these cases, compared to cases where only the turbine spacing is varied.

Overall, the proposed novel modeling method demonstrates exceptional accuracy in interpolating new wake flow fields from limited data, thereby alleviating the challenges associated with acquiring wind farm wake flow field data.

In addition to requiring minimal input information, the proposed method also exhibits high computational efficiency. To evaluate its performance, we compared the runtime of our method with that of LES simulations and commercial software for similar cases. The comparison results are summarized in Table \ref{table_efficiency}. 
Due to the inherent complexity of its physical modeling, LES demands the greatest computational resources and incurs the longest runtime among the evaluated methods. In contrast, the commercial software Meteodyn WT integrates RANS with the Jensen wake model to simplify the physical modeling process, thereby achieving a substantial improvement in computational efficiency. Although the proposed method does not exhibit a significant advantage over Meteodyn WT in generating full three-dimensional data, it provides the flexibility to generate wake field data at specified heights without the need to resolve the entire three-dimensional domain. As a result, two-dimensional wake field data can be generated within two minutes using the proposed method.

\begin{table}[H]
\centering
\caption{Comparison of methods by dimension, grid size, CPU cores, and computation time}
\begin{tabular}{lcccc}
\hline
Method & Dimension & Grid Points & CPUs & Time (h) \\
\hline
LES & 3D & $3 \times 10^7$ & 480 & 24 \\
WT & 3D & $2 \times 10^7$ & 1 & 1.5 \\
Proposed & 3D & $3 \times 10^7$ & 1 & 3 \\
Proposed & 2D & $10^5$ & 1 & 0.03 \\
\hline
\end{tabular}
\label{table_efficiency}
\end{table}

 This efficiency allows for the rapid generation of large volumes of wind farm wake flow field data, thereby providing substantial data for optimizing wind turbine layout and design in wind farms.

\section{Conclusions}
Wind farm wake significantly impacts power production and turbine lifetime. Analyzing wake-affected flow field data provides a deeper understanding of wake development, which facilitates the design of turbine layouts and control strategies. However, current methods for obtaining wake flow field data face several challenges: observational methods incur high equipment costs, physics-based methods and numerical simulations require substantial computational resources, and AI-based methods necessitate extensive training datasets. To address these issues, this paper proposes a data-driven modeling method for rapid wake flow generation. By extracting and matching similar features from wake flow field data and interpolating between these matched features, the method can quickly and accurately generate new wake flow field data from minimal input. The proposed method was validated in six representative cases, encompassing variations in turbine spacing, turbine size, combined spacing and size, turbine number, and small and large wind direction misalignment. 
Evaluation using the MAPE metric shows errors of less than 1.41\% for the x-y plane, and less than 2.28\% for the x-z plane. The proposed method ensures precision and operational efficiency, making it suitable for cost-effective applications in wind turbine layout and design optimization.

Future research will focus on further evaluating the proposed method under a wider range of conditions, including various turbulence intensities and terrain complexities, to comprehensively assess its accuracy and applicability. Specifically, we plan to test its interpolation performance for different turbine spacings, sizes, and wind direction misalignments across diverse terrain and turbulence conditions. In addition, the stability and robustness of the proposed method will be examined under different grid resolutions and noise levels. Validation tests using real measurement data will also be conducted to demonstrate its practical performance in field applications.

\section*{Acknowledgments}
The authors would like to acknowledge the support from the National Natural Science Foundation of China (Grant Nos. 92152109, 12202059, 12588201, 12202453). B. Carmo thanks the Brazilian National Council for Scientific and Technological Development (CNPq) for financial support in the form of productivity grant number 314221/2021-2.





\bibliographystyle{elsarticle-num-names} 
\bibliography{Manuscript}






\end{document}